\documentclass{article}
\usepackage{graphicx} 
\usepackage[utf8]{inputenc}
\usepackage[margin=1.5cm]{geometry}
\usepackage{titlesec}
\usepackage{tabu}
\usepackage{enumitem}
\usepackage{amssymb}
\usepackage{amsmath}
\usepackage[dvipsnames]{xcolor}
\usepackage{blindtext}
\usepackage{hyperref}
\usepackage{lmodern}
\usepackage{caption}
\usepackage{subcaption}
\usepackage{multirow}
\usepackage{xcolor}
\usepackage{array}
\usepackage{booktabs}
\usepackage{array}
\usepackage{soul}
\usepackage{indentfirst}
\newlist{selectlist}{itemize}{2}
\setlist[selectlist]{label=$\square$,leftmargin=*,noitemsep,topsep=0pt}

\title{An AI Theory of Mind Can Enhance Our Collective Intelligence}
\author{Michael S. Harr\'e, Catherine Drysdale, Jaime Ruiz-Serra}

\begin{document}

\maketitle

\begin{abstract}
    {\it Collective intelligence} plays a central role in many fields, from economics and evolutionary theory to neural networks and eusocial insects, and is also core to work on {\it emergence} and {\it self-organisation} in complex-systems theory. However, in human collective intelligence there is still much to understand about how specific psychological processes at the individual level give rise to self-organised structures at the social level. Psychological factors have so far played a minor role in collective-intelligence studies because the principles are often general and applicable to agents without sophisticated psychologies. We emphasise, with examples from other complex adaptive systems, the broad applicability of collective-intelligence {\it principles}, while noting that {\it mechanisms} and {\it time scales} differ markedly between cases. We review evidence that flexible collective intelligence in human social settings is improved by a particular cognitive tool: our Theory of Mind. We then hypothesise that AIs equipped with a theory of mind will enhance collective intelligence in ways similar to human contributions. To make this case, we step back from the algorithmic basis of AI psychology and consider the large-scale impact AI can have as agential actors in a “social ecology” rather than as mere technological tools. We identify several key characteristics of psychologically mediated collective intelligence and show that the development of a Theory of Mind is crucial in distinguishing human social collective intelligence from more general forms. Finally, we illustrate how individuals, human or otherwise, integrate within a collective not by being genetically or algorithmically programmed, but by growing and adapting into the socio-cognitive niche they occupy. AI can likewise inhabit one or multiple such niches, facilitated by a Theory of Mind.
\end{abstract}

\section{Introduction \label{Introduction}}

{\it All intelligence is collective intelligence.}~\cite{levin2023bioelectric} \\

Collectives are capable of achieving things that individuals alone cannot. Notwithstanding the simplicity or complexity of the individuals, their aggregate behaviour can often be understood as a complex processing of information that individuals store, modify, and transfer between each other producing `useful' collective behaviour at the scale of the whole collective. In most instances of {\it Collective Intelligence} (CI), where the agents might be ants in an ant colony, bees in a beehive, or neurons in a neural network, the individual is not aware of the drivers of their behaviour or the behaviour of other agents. For example, a single neuron is neither aware of its own internal processes nor that of a neuron it is connected to, nor is it aware of the end goal to which its activity contributes. Despite both this lack of awareness and the lack of a centralised controller, evolutionary and learning processes have produced an intricate, precise, and highly adaptive system that is capable of functional behaviour that would be impossible for any single neuron to achieve. In other instances of CI, such as teams of humans, or businesses interacting in economic markets, the agents themselves may be highly complex and exhibit varying degrees of purposefulness and awareness. Within this context, we draw attention to the role of psychological factors in improving the CI of human social collectives and quantifying the intelligence of social collectives, both natural and artificial. In this Introduction we review key aspects of collective intelligence and recent progress in placing AI into the collective intelligence framework.

\subsection{Individual agents and their collective intelligence}

In order to understand how collectives process information, we first consider the variety of ways in which agents interact. The topology of the network describing agent-to-agent interactions is well known to be important for the proper functioning of social groups~\cite{momennejad2022collective,migliano2022origins}. In particular it has been shown that mammalian social groups exhibit patterns of fractal-like topologies~\cite{hamilton2007complex,hill2008network} that are a result of a cognitive ability to form discrete social connections between conspicifics~\cite{harre2016social}. These links are often both spatially and temporally transient; people meet for a while, go their separate ways, and come back together later. Despite this transience, individual connections are often the basis of long term social relationships between specific individuals as in pair-bonding and friendships. Consequently an important distinction can be made regarding connections between agents in complex adaptive systems: they can be more fluid-like or more solid-like~\cite{sole2019liquid}. For example the links between neurons in the brain are relatively fixed in nature when compared to the brief communicative interactions between ants, either instantaneous interactions between individual ants or via transient pheromone trails that coordinate the behaviour of large numbers of ants. Sol\'e and colleagues~\cite{sole2019liquid,pinero2019statistical} identify a distinction between solid brains, in which interactions between agents fixed in place are highly persistent in time (e.g. neural networks, spin glasses) and liquid brains, in which interactions between highly mobile agents are much more short-lived (e.g. ants, immune cells). As Sol\'e \textit{et al.} note regarding liquid brains~\cite{sole2019liquid}: ``Here there are no neural-like elements and yet in many ways these systems solve complex problems, exhibit learning and memory, and make decisions in response to environmental conditions.''

All biological agents are composed of sub-units such as organs, cells, and molecular networks~\cite{levin2019computational,levin2022technological,levin2023darwin}. Cells in particular are the simplest living organisms with individual intelligence, or {\it competencies}~\cite{levin2019computational,fields2022competency}, within their native contexts. Here, we briefly focus on the archetypal single-cell intelligence, the neural cells. It is well understood that the central nervous system is a highly developed, adaptive, complex system that exhibits emergent computational characteristics~\cite{hopfield1982neural}, both in biological and artificial neural networks. Naturally the artificial models are simplifications but the extent to which they are simplifications is not so well understood. In a 2021 study, Beniaguev \textit{et al.}~\cite{beniaguev2021single} concluded that between five and eight layers of an artificial deep neural network are required to approximate the input--output mapping of a (single) cortical neuron and that the dendritic branches can be understood as spatiotemporal pattern detectors. This demonstrates that a single neural cell can be modelled as an artificial agent with highly complex computational capabilities situated within an adaptive, complex network of other highly complex agents, all signalling to one another. These results can be compared with earlier studies in which neurons were modelled as a Bayesian agent that is trying to infer the state of a hidden variable~\cite{deneve2008bayesian}. In each of these interpretations, a single cell can be seen as an agent with computational competencies situated within the context of a network that is slowly and adaptively changing around it. 

We can also compare the competencies of neural cells in networks to the individual competencies of ants in an ant colony. In a recent study~\cite{kay2024ant} it was shown that social structures of some ant colonies are conserved between species that are separated by more than 100 million years of evolution. In the five species studied by Kay \textit{et al.}~\cite{kay2024ant}, they found two social clusters and similarities in the division of labour that are preserved between the species. In a different study, Richardson \textit{et al.}~\cite{richardson2021leadership} showed that individuals within an ant colony play an important {\it leadership} role and that the behaviour of these individuals significantly improved the collective performance of the ants. Ants are also capable of changing their social structure in the event of pathogenic infestation of their colony. In a 2021 article, Stockmaier and colleagues~\cite{stockmaier2021infectious} review the research on {\it social distancing} and other social restructuring that occurs with conspecifics in order to reduce the impact of pathogens by changing their social cues, signals, and other behaviours for the collective benefit of the colony. These two very different systems, neural networks and ant colonies, are examples of complex collective intelligences where the individuals (neurons, ants) are complex in their own right, but they signal each other in order to restructure their relationships so as to adapt their collective competencies to external signals. The neural networks are prototypical {\it solid brains} and ant colonies are prototypical {\it liquid brains} and there has been recent progress in developing collective intelligence in the context of collective adaptation~\cite{galesic2023beyond}.

Human social interactions can also be viewed as a form of liquid intelligence. Migliano \textit{et al.}~\cite{migliano2022origins} discuss the `fluidity' of social relations in early human societies: ``Quantification and mapping of hunter–gatherers' social networks has revealed details of a fluid and multilevel sociality, where friendship links connect unrelated mobile households into camps of temporary composition''. They describe the key characteristics of early human society, such as egalitarianism, division of labour, cooperative living with unrelated individuals, multi-locality, fluid social structures, and high mobility between campsites, which might be  thought of as a liquid brain composed of social interactions that both cluster and disperse in order to store, modify, and transfer information via social networks. The notion that human social interaction might be a form of computation is not new: Mirowski, Axtell and colleagues~\cite{mirowski1998markets,l2003economics,mirowski2007markets} have suggested that economic markets are a form of computation by which prices can be derived, and Harr\'e recently hypothesised~\cite{harre2022entropy} that this could be measured using information theory as had been done earlier for financial markets~\cite{harre2009phase,harre2015entropy}. As Axtell \textit{et al.}~\cite{l2003economics} wrote: ``There is a close connection between agent computing in the positive social sciences and distributed computation in computer science, in which individual processors have heterogeneous information that they compute with and then communicate to other processors.''

The emergence of computation in multi-agent systems is a well-studied area of complex adaptive systems~\cite{langton1990computation,moore2018inform}. For example neuroscience has used information theory to describe the storage, transfer, and modification of bits of information in biological neural processes~\cite{wibral2014directed}. More broadly, Integrated Information Theory (IIT)~\cite{tononi2016integrated,mediano2022integrated} has been put forward as a measure of the emergence of `consciousness' in generic (non-biological, non-neural) systems. In this case, some forms of IIT explicitly use information theory~\cite{barrett2011practical,mediano2018measuring} to measure the amount of non-trivial computation a system is carrying out. More generally, there is a move towards understanding complex adaptive systems in computational terms~\cite{prokopenko2009information,lizier2012local} by empirically measuring the inter-agent flow of information~\cite{bossomaier2016transfer}. \\

\subsection{Recent developments using AI agents in complex adaptive systems}

At the meso-scopic level, between artificial neurons and the collective behaviour of AI, is the emerging discpline of {\it machine behaviour}~\cite{rahwan2019machine}. This field studies intelligent machines not as engineering artefacts but as a distinct class of actors with unique behavioural patterns and ecological dynamics~\cite{johnson2013abrupt}. Motivated by the increasingly pervasive integration of algorithms into human society and the challenges in formalising and predicting their complex effects, this field examines machine behaviours empirically, similar to how ethology and behavioural ecology study animals by integrating intrinsic and environmental influences~\cite{aharony2011social}. In this context, it bridges micro- and macro-level questions about algorithmic interactions and their societal implications, using methods like randomised experiments and observational studies from the behavioural sciences. By drawing parallels with biological frameworks, including Tinbergen’s four dimensions of function, mechanism, development, and evolution~\cite{tinbergen2005aims}, machine behaviour explores how machines interact with their environments and stakeholders, creating novel trajectories of evolution and influence that differ from organic systems. This interdisciplinary approach highlights the necessity of understanding AI's dual role in shaping and being shaped by human systems, suggesting careful examination of its impact to harness its benefits while mitigating risks in a socio-technical context.

We can illustrate the current state of specific AI frameworks in complex adaptive systems using Large Language Models (LLMs) as an example~\cite{burton2024large}. LLMs are powerful tools for enhancing CI, but they do not yet function as autonomous agents in their own right, but instead act as mechanisms that augment human capabilities in collaboration, creativity, and decision-making. Their potential applications are highly diverse, including increasing accessibility and inclusion in online collaborations, accelerating idea generation, mediating deliberative processes, and aggregating information across groups. However, their use also introduces risks, such as disincentivising contributions to collective knowledge, fostering illusions of consensus, reducing functional diversity, and enabling the seamless production of false or misleading information. Viewed with a wide lens, LLMs not only support CI but are themselves products of it, having been trained on collective human data and refined through collective feedback. This places the use of LLMs in the context of a tool that can improve pre-existing human CI, but the LLM itself does not have agential properties in the sense that its behaviour is not produced by selecting from multiple, internally developed goals, it has very little autonomy, and it is not typically connected to a real-time working environment. 

In contrast to LLMs as non-agential AIs, new work is developing cognitively informed approaches to AI-human cooperation in which they have multiple self-regarding goals that account for the goals of others, allowing them to behave with a degree of cooperative agency. In recent work by Crandall and colleagues~\cite{crandall2018cooperating}, it was shown that general human–machine cooperation is achievable through a sophisticated yet fundamentally simple set of algorithmic mechanisms, with results showing that an algorithm can cooperate with humans and other machines at levels comparable to human–human cooperation in repeated stochastic games (RSGs). Through a comparison of existing algorithms, the development of a novel learning algorithm that integrates high-performing strategies with human-conducive signaling mechanisms, the algorithm (S\#) was shown to form and maintain cooperative relationships across diverse RSGs. Key to its success is the inclusion of `cheap talk'~\cite{sally1995conversation} which significantly increases mutual cooperation and highlights the importance of signaling mechanisms that are aligned with human understanding. S\# also employs a variety of expert strategies that can be adapted to different partners and game types. Its expert-selection mechanism, modelled on the psychologically based recognition-primed decision-making~\cite{klein1993recognition}, allows S\# to achieve a level of flexibility and generality unmatched by traditional expert-selection methods, illustrating its potential for fostering effective cooperation in complex, real-world interactions.

Similar results have been achieved in games with more realistic complexities, approaching human levels of subtlety in social reasoning within multi-agent, mixed human-AI, environments. Work carried out on an AI playing the game {\it Diplomacy}~\cite{meta2022human} has shown that AIs can successfully defeat expert human players in complex social games of cooperation and deception without being recognised as an AI. The AI, called Cicero, integrates dialogue generation, strategic reasoning, and message filtering into its decision-making architecture in order to pass messages to other players (all of them humans who were unaware Cicero is an AI) in order to strategise about future plans in the game. Its dialogue system is based on a pre-trained language model that is further developed on human game dialogue, grounded in both the dialogue history and the game state. Dialogue generation is controlled through `intents'---planned actions involving Cicero and the person it is messaging---enhancing contextual relevance and strategic alignment while offloading the responsibility of learning game legality and strategy to other modules. Cicero's strategic reasoning module employs a planning algorithm that predicts opponents' policies based on game state and dialogue, selecting optimal actions guided by reinforcement learning with constraints to align with human-like behaviour. The approach used by Cicero has been described as an elementary form of modelling the cognitive state of other players, i.e. having a Theory of Mind. Messages are dynamically generated and filtered to ensure they are coherent, consistent with intents, and strategically sound, allowing Cicero to integrate negotiation and strategic planning in a manner indistinguishable from human players. As one of the researchers describes it~\cite{gent2023machine}: ``I think that the substance of what we're doing is understanding the beliefs, goals and intentions of the other players.''. Both Cicero and S\# illustrate that language combined with psychological aspects of reasoning and a strategic awareness of the environment are able to build coordination and cooperation with humans well enough for the AI to pass as a human. In order to move towards AI based more closely on human psychology~\cite{gigerenzer2024psychological} to enhance our collective intelligence~\cite{cui2024ai} requires a greater awareness of the ecology and sociology of AI~\cite{tsvetkova2024new}, and the development of interdisciplinary approaches to understanding the macroscopic consequences of these technologies.

In this article we will use information theory to quantify how much computation in a CI is `emergent' and how much is simply independent information processing by single agents. In general, we wish to capture the notion of the whole (computational process) being greater than the sum of the (independent) parts. We translate this to the simple notion that to the extent to which this inequality holds: \texttt{Whole - }$\sum($\texttt{Parts}$) > 0$ is the extent to which we will say a system exhibits non-trivial CI, noting that there are multiple possible implementations of this approach~\cite{kanwal2017comparing}. The \texttt{Parts} is how much computation a single agent is carrying out from one time step to the next such that the sum is the total of all agents' independent computations. The \texttt{Whole} is the totality of computation in the system, it includes all single agent computations, pairwise computations, and higher order interactions between agents. Our measure will not be unique in any of its specifics, but it serves to quantify the CI of a system for comparative analysis. This approach also has much in common with that of Moore \textit{et al.}~\cite{moore2018inform} in which information theory is used to measure the collective intelligence in biological systems.

Not only is there diversity in the types of systems that can show positive measures of CI, but the ways in which agents manipulate a system's computations is diverse as well. Take for example Watson and Levin's discussion of a scientist manipulating the intercellular signalling in order to change their collective outcome~\cite{watson2023collective}: \\
\begin{quote}
    This framework [of collective cellular intelligence] makes a strong prediction: if intercellular signalling (not genes) is the cognitive medium of a morphogenetic individual, it should be possible to exploit the tools of behavioural and neuro-science and learn to read, interpret and re-write its information content in a way that allows predictive control over its behaviour (in this case, growth and form) without genetic changes. 
\end{quote}

\noindent A counter question is: How can single agents, such as human leaders, have predictive control over a social group? Just as a scientist \textit{external} to a cell collective can manipulate inter-cellular signalling to control the outcomes of the cell collective, a leader \textit{internal} to a human collective can manipulate inter-personal behaviours to control the outcomes of the human collective. In both cases, an agent with a goal-directed psychology is acting on inter-agent relationships, i.e. inter-cellular or inter-personal, to control outcomes at the next level higher, i.e. organism-scale or societal-scale.

\subsection{The approach taken in this article}

In this article we will reframe the question Watson and Levin asked in the following way: \textit{What is there in human psychology that allows us to learn to read, interpret, and re-write our interpersonal information content in a way that allows predictive control over our collective behaviour?} and consider the answer in the context of pervasive, agential AI. We review some of the extensive literature showing that our Theory of Mind (ToM) is a suite of cognitive skills that allows individuals to have goal directed control over collective outcomes. Originally ToM was used to describe our ability to infer the unobserved mental states of other people~\cite{frith2010social} such as desires and beliefs, an ability humans are particularly good at and other animals much less so~\cite{penn2007lack,krupenye2019theory}. But recently it has been shown that ToM is predictive of group performance as 
well~\cite{woolley2010evidence,engel2014reading}, empirically demonstrating the role of ToM in going beyond representations of the internal states of others to using that knowledge in a social setting to improve the collective outcomes for the group. In order to model ToM in a tractable fashion, we will focus on the narrower {\it game theory of mind}~\cite{yoshida2008game}, and the {\it Beliefs, Preferences, and Constraints} (BPC) interpretation of game-theoretic decisions put forward by Gintis~\cite{gintis2006foundations}. In this approach, what agents understand of other agents' hidden states are the BPC that structure their observable behaviours.

We will consider this question in the framework of agent interactions that extend agent utilities in a simple but novel way. We quantify our results using information theory to show the impact that a well-developed ToM has to direct agents' behaviours in order to increase our CI. The models are simple but they illustrate the central notion that understanding the ``beliefs, preferences, and constraints''~\cite{gintis2007framework,gintis2007unifying} of others can be used to improve the CI of a complex social system. The wider purpose of this work is to place recent developments, such as human-AI co-evolution~\cite{pedreschi2024human}, AI-enhanced collective intelligence~\cite{cui2024ai}, the collective sociology of AI and humans~\cite{tsvetkova2024new}, the connection between ToM and socio-cultural niche evolution~\cite{veissiere2020thinking}, and the design of intelligent cyber-physical ecosystems~\cite{friston2024designing}, into the context of measuring hybrid CI, CI facilitated human sociology via ToM, and how concepts from ecology, psychology, and economics can help us build a better understanding of the joint future evolution of humans and AI.

In Section~\ref{sec:topo-structure}, we describe the liquid--solid spectrum of interacting agents, review existing models of ToM, and provide perspective on the interplay between social network structures and ToM. We specifically draw attention to three key ideas that provide empirical support for either ToM or CI at the scale of human collective behaviour. First, the liquid--solid brain hypothesis establishes a cognitive spectrum from rigid and persistent to fluid and dynamical interactions between agents that exhibit collective intelligence. This approach is independent of the spatial scale of the collective, for example it can be used to study neurons, insect colonies, or human social interactions. Second, we consider the evidence for ToM, its structure at the individual agent level as a method for agents to model the unobserved control parameters that influence other agents' behaviour. We then show how our use of language, a fluid communication channel that temporarily couples agents together in order to share information and modify behaviour, is tightly connected with our development and use of ToM. Finally, we consider the role of network topology in CI and in particular how recent work has measured the CI of social groups and related it to the individual's capacity for ToM. In Section~\ref{sec:illustrative-examples}, we provide illustrative examples supporting different aspects of our argument, introducing our measure of computation and applying it to a simple empirical example. Here we note that the effectiveness of a collective depends on the macroscopic structure of interactions between agents such that any analysis will need to account for more than simple dyadic interactions and so we introduce these higher order interactions. In computing the CI between agents, a computer interacting with a monkey, we illustrate a game theory environment with an AI-biological hybrid system in which the CI value can be measured using observable behaviours. In Section~\ref{4.0} we review the psychology of social fluidity and the variety of social outcomes that this fluidity makes possible. We also use a simple multi-agent system to describe how a ToM can be used to improve the computational processes, i.e. the CI, of interacting agents. It is here that we connect the hidden variables (i.e. meta-parameters) argument of ToM discussed earlier with the parameters of game theory describing the beliefs, preferences and constraints that incentivise decisions~\cite{harre2023testing,harre2022can}. We also provide a formal interpretation of the higher order interactions introduced earlier in terms of game theory. Finally, in Section~\ref{5.0}, we discuss the broader implications of this approach.\\

\section{Cognitive morphospaces: The network topology of agents' interactions}\label{sec:topo-structure}

The emergence of cognitive networks marked a pivotal moment in our evolutionary history~\cite{WoodRachel2019Iroe}. Earlier, microorganisms had developed collective structures capable of responding to the physical environment, particularly conditions threatening individual cells~\cite{BaluškaFrantišek2016OHNH} and survival became dependent on information exchanges within these groups. Habituation and the ability to minimise the energy response to danger stimuli is considered one of the simplest forms of learning and has been extensively studied in simple collective intelligences such as slime moulds~\cite{BoussardA2019Miap,ThompsonRichardF.2009HAh}. Similarly bacteria use quorum sensing to coordinate their behaviour based on the density of the population of their community to coordinate responses, leading to a change in gene expression and function regulation e.g. bioluminescence, release of toxins, and biofilm formation~\cite{LiZhi2012QsHb}. Information processing and problem solving capabilities developed using a variety of interaction types and network topologies long before the appearance of central nervous systems with fixed neuronal structures~\cite{BaluškaFrantišek2016OHNH}. But this leads to an interesting question regarding the {\it typologies} of agent-to-agent interactions and the intelligence these structures might enable, a question that can be approached by looking at the {\it morphospace of collective intelligence}.

A morphospace is a theoretical framework used to simplify and organize the complex shapes and forms of organisms, typically focusing on external anatomical features, into a more manageable space representing their potential variations. For instance, in~\cite{avena2015network}, a three-dimensional morphospace was constructed for organisms with shells, where the diversity of shell shapes is described by three key parameters: a deviation angle, a translation factor, and a growth factor. This reduces a high-dimensional structural space to a lower-dimensional one, where a small number of parameterised properties captures key variations between forms. Morphospaces have found uses in many different fields of research, including the body shapes of fish~\cite{claverie2014morphospace}, network topologies~\cite{avena2015network}, and the structural forms of language~\cite{seoane2018morphospace}. 

These structures, which represent recurring patterns of trait variation, are of great interest to evolutionary biologists because they may indicate shared evolutionary processes and their constraints. They are also of great interest to researchers investigating the underlying structures of collective intelligence~\cite{levin2022collective}, as intelligence in its different forms may also be subject to shared evolutionary mechanisms and constraints. With this in mind, work has been done in studying a variety of morphospaces related to collective intelligence and in the next section we review some examples.

\subsection{Network topology and the ``Solid Brain, Liquid Brain'' framework}
\label{sec:solid_liquid}

In the most general of terms, a cognitive network has multiple information processing components that exchange information with each other and interact with their environmental context. Sol\'e {\it et al.}~\cite{sole2019liquid} identified two key dimensions for categorising different types of cognitive networks: the system's physical state---either more {\it liquid} or more {\it solid} in nature, differentiated by how freely individual agents (components) can move in space---and the presence or absence of neurons. The collective dynamics of a large population of agents is influenced by the individuals' mobility which dictates how they respond to signals both internal and external to the collective. To help conceptualise this diversity, Sol\'e \textit{et al.}~\cite{sole2019liquid} developed a morphospace and taxonomy in order to compare and contrast the physical states of different types of CI. In this way they were able to consider the physical properties that form constraints on the computations achievable by a system. This then poses an interesting question: Is the entire space of possible {\it intelligences} being exploited by either synthetic biology or abiotic computation?

Liquid brains exhibit cognitive behaviours without neurons. For example, models by Watson and colleagues~\cite{Watson2011GlobalAdaptation, Watson2011OptimizationSelfmodeling} illustrated how systems of self-interested agents, driven by a simple mechanism of strengthening beneficial connections, can lead to robust group-level adaptation and problem-solving. 
This self-organisation, akin to Hebb's rule in neural networks, enables the system to recall and consequently leverage past configurations that were successful. This also allows the system to generalise from experience and then predict beneficial states it has not encountered before, highlighting how decentralised actions can produce a form of CI that guides the system towards greater global utility. The agents in these models are very simple, but this need not be the case---each agent within a collective may itself have a solid brain as in human social networks where fluid social interactions allow each solid brain to connect and communicate with other solid brains for the benefit of the collective~\cite{brede2022sensing}. This is extended to a hybrid model by Kao \textit{et al.}~\cite{KaoAlbertB2019Mswg}, wherein the modular organisation of mobile animal populations (as an example of a liquid brain) suggests that certain communication pathways exhibit localised and persistent characteristics, akin to those in solid brains. 
These pathways enhance collective decision-making in complex environments and in turn raise important questions regarding the relative strengths of liquid and solid brains in different contexts. In particular, the conditions under which liquid brains outperform solid brains, especially in terms of adaptability and scalability, remain an open area of investigation. Which computational problems are more effectively solved by liquid brains vis-\`a-vis solid brains remains to be better understood.

Whole classes of models that describe which biological or artificial structures are capable of some form of computation, either in potentia or in practice, can often be usefully represented using morphospaces~\cite{budd2021morphospace}. Computational morphospaces~\cite{sole2019liquid} have proven effective in studying key properties of complex adaptive systems, for example the statistical mechanics of information processing and structural variations~\cite{arsiwalla2017morphospace,bahri2020statistical}.
This sheds light on how energy constraints influence the evolution and adaptability of neural networks across a variety of biological systems.  Arsiwalla \textit{et al.}~\cite{arsiwalla2017morphospace} have examined how liquid brains and solid brains fit within such a framework, by comparing the flexibility and adaptability of different neural architectures~\cite{arsiwalla2017morphospace}.  The dimensions of their framework are three different types of complexity that a system may display: autonomic, computational, and social complexity. We suggest that other axes for consideration are a system's solid--liquid dimension as Sol\'e {\it et al.}~\cite{sole2019liquid} and Oll\'e-Vila {\it et al.}~\cite{olle2016morphospace} have done, as well as the system's degree of ToM (see Section~\ref{2.2}) and the system's information processing capacity (see Section~\ref{sec:illustrative-examples}). 

We posit that the collective intelligence that emerges from liquid brains (human social networks) is enhanced by our individual capacity for a ToM, where individuals are aware of the goals of others as well as that of the collective, which can then be achieved by adaptation at the local level. Rather than collective intelligence arising as an epiphenomenon or byproduct of agents interacting, our ToM allows agents to causally affect the outcome of the system they are a part of. 


\subsection{Models of another agent's internal states \label{2.2}}

Frith and Frith defined Theory of Mind (ToM) as how we explain other people's behaviour on the basis of their internal cognitive states, i.e. their knowledge, beliefs, and desires~\cite{frith2010social}. There is now a vast literature on this topic in psychology, sociology, and more recently artificial intelligence, but for the purposes of this article we restrict ToM to apply to the subset of the beliefs, preferences, and constraints of other agents in the sense of incentivised decisions. This borrows from the BPC model~\cite{gintis2006foundations} put forward as an approach to understanding the socio-cognitive aspects of human decision-making~\cite{gintis2007unifying}.

One way to interpret the BPC model is that it imposes structural constraints on the process by which decisions are made, and then optimal decisions are discovered within these constraints by parametric variation. Recent work by Peterson \textit{et al.}~\cite{peterson2021using} compared more than 20 structurally constrained models of individual decision-making using human data. That study was extended to human data during strategic interactions by Harr\'e and El-Tarifi~\cite{harre2023testing} in order to test agents' constrained representations of other agents. This extends Yoshida \textit{et al.}'s~\cite{yoshida2008game} notion of a {\it Game Theory of Mind} to a larger variety of models in which an agent's strategic reasoning about other agents modulates their behaviour. 

In the field of artificial intelligence, Jara-Ettinger~\cite{jara2019theory} proposed the use of Inverse Reinforcement Learning (IRL) as a model of agential ToM, 
whereby agents modelling other agents' mental states is equivalent to inferring an unobserved world model the other agent uses in their decision-making, as well as their reward function. 
Jara-Ettinger discusses some key limitations of IRL, such as the difficulty in recovering an agent's beliefs and desires even while assuming that all agents are identical in their choice-making.
IRL has been implemented in many different algorithmic forms, and their applicability as a basis for ToM was recently reviewed by Ruiz-Serra and Harr{\'e}~\cite{ruiz2023inverse} and recent work by others in developing ToM for artificial intelligence in collaboration with people~\cite{nguyen2022theory,zhao2023teaching,gupta2023fostering}.

What is missing in the Jara-Ettinger perspective is how different internal models, specifically different drivers of behaviour predicated on the BPC of others, influence how agents interact with one another in a social network. Critically, people use and improve upon these information carrying interactions in their social networks, as discussed next. This requires agents to have more than a representation of the world model used by other agents, it needs to be a {\it social} world model of how agents are influenced by their interactions with other agents. Beyond inferring internal states, Shteynberg \textit{et al.}~\cite{shteynberg2023theory} distinguish between awareness of the self, awareness of the self in relation to others, and an awareness of the collective, which is more than just the sum of the self and others, a type of collective awareness they have called {\it Theory of Collective Mind}. In a similar vein, Shum \textit{et al.}~\cite{shum2019theory} proposed a generative model for understanding multi-agent interactions called Composable Team Hierarchies. This approach used stochastic games in conjunction with multi-agent reinforcement learning in order to infer relationships between agents and predict future behaviours. 

In cognitive science, recent progress has been made in identifying levels of ToM ability and placing them in a hierarchical structure (see also Yoshida \textit{et al.}'s Game Theory of Mind~\cite{yoshida2008game} and a recent review by Harr\'e~\cite{harre2022can}). 
A cognitive agent with \textit{zeroth order} ToM attributes no cognitive ability to other agents, whereas \textit{first order} ToM attributes some cognitive abilities to others, and so on. 
Here we summarise the orders as described by Lombard and G\"ardenfors~\cite{lombard2023causal} and use  {\it A} and {\it B} to identify two agents that may or may not have any cognitive ability:
\begin{itemize}
    \item Zero order ToM: Both {\it A} and {\it B}'s behaviour is governed by instinct, reflexes, and conditioning and so direct perception of the agents' interactions with their environment and each other is all that is needed to understand their behaviour.
    \item First order ToM: {\it A} and {\it B} can be attributed with emotions, attention, desires, intentions, or beliefs, but neither agent attributes these properties to any other agent, including themselves.
    \item Second order ToM: {\it A} attributes to {\it B} internal cognitive states, and that {\it A} uses this knowledge to understand {\it B}'s behaviour. This is the lowest level at which {\it A} attributes hidden (cognitive) variables to {\it B} in order to explain the causes of {\it B}'s actions, i.e. it abstracts causation away from direct perception of the causes of behaviour.
    \item Third order ToM: {\it A} attributes to {\it B} an understating of {\it A}'s internal states. To borrow an example from game theory (stag hunt) and early human society (hunter gatherers), when a hunting party stalks an animal everyone shares a common goal whereby each person {\it A} knows that the others are aware of {\it A}'s goal, such that this cognitive state will causally inform {\it A}'s actions. Lombard and G\"ardenfors note that it has not been conclusively demonstrated in nonhuman primates and indicate there are alternative views~\cite{haidle2010working,lang2012pleistocene}.
    \item Fourth order and higher ToM: {\it A} has an awareness of at least two mental states, their own and that of {\it B}. For example Happ{\'e} reviewed the evidence and suggested that reflecting on one's own cognitive state relies on the same neuro-psychological functions as those we use to attribute thoughts to others~\cite{happe2003theory}.
\end{itemize}

Most interactions between agents across all species will be of the zeroth or first order ToM, where neither agent has higher order cognitive states, has no sense of being aware of other agents nor any self-awareness. There is no sharp line that unambiguously distinguishes between stimulus-response, conscious, and self-conscious agents as it is an open area of research and is a multi-dimensional phenomenon~\cite{birch2020dimensions}, and so the conscious status of agents is likely more incremental than the discrete levels of this scale would indicate, but is a useful framing device. 

This leads to another core finding related to ToM: the interplay between language and ToM in humans. Chomsky has noted both how well developed this ability is in humans and how communication sits in relation to our mental states~\cite[p. 10]{chomsky2007biolinguistic}:
\begin{quote}
Communication is not a matter of producing some mind-external entity that the hearer picks out of the world, the way a natural scientist could. Rather, communication is a more-or-less affair, in which the speaker produces external events and hearers seek to match them as best they can to their own internal resources. Words and concepts appear to be similar in this regard, even the simplest of them. Communication relies on largely shared cognoscitive powers, and succeeds insofar as similar mental constructs, background, concerns, and presuppositions allow for similar perspectives to be reached. If that is true---and the evidence seems overwhelming---then natural language diverges sharply in these elementary respects from animal communication.
\end{quote}
That is to say, effective communication between people requires an encoding-transmission step and a reception-decoding step, both premised on shared cognitive representations in order to be understood, which we contend is made possible via ToM, i.e. a (shared) representation of each other's hidden cognitive variables.

Recent work has informed Chomsky's view here, linking language and ToM with our ability to carry out causal reasoning, particularly in social contexts. In Lombard and G\"ardenfors they proposed three key hypotheses relating ToM to cognitive structures~\cite{lombard2023causal} (quoted, emphasis added): 
\begin{itemize}
    \item Theory of mind is an integral element of {\it causal cognition}; 
    \item Generally speaking, the more advanced causal cognition is, the more it is dependent on theory of mind; and \item The evolution of causal cognition depends more and more on {\it mental representations of hidden variables}. [...] [C]ausal cognition allows us to reason from a network of hidden variables ...
\end{itemize}
Language use is related to this causal reasoning about hidden variables via the representational view of language due to their intersection with ToM~\cite{farrar2002early}. In the representational view of language people use specific grammatical structures to represent complex events and then to reason from them~\cite{de2021role}. These structures serve as a cognitive tool, particularly in representing others' mental states, enabling the expression of false beliefs, lies, or mistakes. It has also been shown that these are strong predictors of children's false belief understanding, a canonical test of ToM, and targeted syntactical training can improve false belief reasoning~\cite{de2002complements}. Notably, in developmental learning, language has a stronger influence on ToM abilities than vice versa~\cite{milligan2007language}. From this we identify causal reasoning about hidden cognitive states with causal reasoning about the behaviour of others via a shared representation of our social environment, mediated by specific syntactical structures. 

How is language `causal' though? That is, how do cognitive representations causally manifest themselves in collective outcomes via language? To some extent this might be a natural assumption to make, but its significance has been shown experimentally in that the words we choose, that grammatically represent our internal models of the world, have consequences in policymaking. In work by Thibodeau and Boroditsky~\cite{thibodeau2011metaphors} they have shown that two different metaphors used to describe the same numerical data regarding crime in a city: as either an infectious  disease or a rampaging beast, causally influenced the likelihood of policies people chose: either preventive or punitive measures. This, they have argued~\cite{thibodeau2017linguistic}, is the conceptual scaffolding through which we reason and that these linguistic metaphors guide our thoughts and behaviours. To date there has been little work on the levels of ToM for AI or the causal nature of language models, and this gap in the literature will need to be filled if we wish to understand how AIs can play an effective role in human collective outcomes. But if language is causal, the degree to which our language impacts outcomes is also contingent on the structural properties of the networks over which this information spreads~\cite{bakshy2012role}. We discuss these aspects next in the context of shared cognitive representations.


\subsection{Configurations of our social networks inform individual reasoning}

Frith and Frith have previously considered the benefits that accrue to humans via our ToM~\cite{frith2010social}, but what is functionally happening when we use our ToM in social groups? It has been shown that in early hunter-gatherer societies that some emergent phenomena at the social level, e.g. Dunbar's Number~\cite{dunbar1993coevolution,dunbar2007evolution}, are a consequence of the layered, fractal topology of social networks~\cite{dunbar1995social,hamilton2007complex,hill2008network}, and that these are in turn the product of very specific, discrete, cognitive constraints at the individual level that shaped the structures of early human societies~\cite{harre2016social}. At the individual level, a recent review by Momennejad~\cite{momennejad2022collective} collected the evidence for different social network topologies and how they integrate interpersonal knowledge differently, showing that topology allows social networks to serve a rich variety of collective goals. Momennejad also reviews the neuro-imaging evidence showing that humans neurologically encode these topologies and these encodings are shared across the members of a social group. This was also demonstrated in the work of Lau \textit{et al.}~\cite{lau2018discovering} showing that people are able to integrate information about how agents relate to one another in addition to how they relate to oneself in order to infer social group structures. Lastly, there is evidence for improved collective intelligence when individuals with higher competencies in ToM are present in the group, as shown in the study by Woolley \textit{et al.}~\cite{woolley2010evidence}. It was found that, just as there is for an individual person a measure of general cognitive ability, usually denoted $g$, there is an equivalent measure of collective intelligence, denoted $c$, for a group of people. They noted a key explanatory factor of a group's task performance, as measured by $c$, was the proportion of group members who ranked highly on the {\it Reading the Mind in the Eyes} cognitive test introduced by Baron-Cohen and colleagues~\cite{baron1997another,baron2001reading}, a test used to measure an individual's capacity for ToM. 

A key takeaway from the work of Woolley and colleagues~\cite{woolley2010evidence} is that the $c$ factor is not strongly correlated with either the average or maximum intelligence of the individuals in a group. However, it does correlate well with the average {\it social sensitivity} of its members, the evenness of the distribution of contributions to group discussions, as well as the proportion of people in the group who rate highly on a ToM test. So there is considerable evidence for the role our ToM plays in group performance, how this shapes interpersonal interactions between people, and, as a consequence, the emergent topological properties of our social groups. This provides support for the argument that ToM is (one of) the individual, bottom-up mechanism(s) through which agents form higher-order social structures with measurable collective intelligence. This may be how we ``learn to read, interpret and re-write our interpersonal information content'' as we asked following the Watson-Levin quote in Section~\ref{Introduction}. Next, we illustrate these ideas in a dyadic and then a triadic example of agents interacting with each other and exhibiting non-trivial measure of CI. 

\section{Topological and cognitive structures in CI: illustrative examples}\label{sec:illustrative-examples}

\subsection{A dyadic example of individual learning at short time scales}
To illustrate how we will use information theory as a proxy for Woolley \textit{et al.}'s $c$ intelligence, we use a dataset that was previously studied~\cite{harre2018strategic} to measure the information flow in an iterated economic game experiment between monkeys and computers, based on data from an earlier study by Lee \textit{et al.}~\cite{lee2004reinforcement}. In that study, the experiment had a monkey playing the matching pennies game against a computer algorithm for a reward, the (Nash) optimal reward for the monkey was received if it plays 50:50 across its two choices. A simplified description of the three algorithms used by the computer follows, see Lee \textit{et al.}~\cite{lee2004reinforcement} for the exact descriptions: 
\begin{itemize}
    \item Algorithm 0: Play uniformly and independently of the monkey's choices, \item Algorithm 1: The computer stores the history of the choices made by the monkey, then to predict what the monkey would do in each trial the computer calculates the conditional probability of the monkey's choice given the monkey’s choices in the preceding 4 trials, a Null hypothesis was used to test if the monkey played 50:50 and if it was not rejected the computer plays 50:50, if it was rejected the computer plays probability $1-p$ against the monkey's probability of playing $p$ 
    \item Algorithm 2: Uses the same algorithm as 1) but includes both choices and rewards in the monkey's history, and then both algorithm 1) and 2) were tested against the Null of 50:50 and if the Null was not rejected the computer plays 50:50, otherwise the best reply was played based on the estimated bias of the monkey.
\end{itemize}

The usefulness of this example is fourfold. First, it is simple enough that computations can be tested and evaluated in an illustrative way so that the central information theory measures can be applied. Second, it uses game theory as the foundational mechanism for the interactions that generate the information flow between the agents. Third, it is complex enough to illustrate key elements of CI with only two agents. Finally, it illustrates the general principles that can be applied in evolution, psychology, AI collectives, and human-AI hybrid settings. Our definition of CI is based loosely on the information theory form of Integrated Information Theory (IIT)~\cite{tononi2016integrated}, where we are neutral to the interpretation of IIT as a measure of consciousness, and will not be carrying out any optimisation over binary partitions of state variables, so in that sense this is not the same as IIT, although there are some similarities. Our definition for a system with $n$ agents is (see Section 2.2.3 of~\cite{mediano2018measuring} and for a general introduction see Battencourt~\cite{bettencourt2009rules}):

\begin{equation}
    \phi(X;\tau) \triangleq I(X_{t}; X_{t-\tau}) - \sum_{i=1}^n I(X^i_{t}; X^i_{t-\tau}). \label{coin}
\end{equation}
For example we might have $X_t = \{X^1_t, X^2_t\}$ as the joint stochastic variable of agents 1 and 2 such as the monkey and the computer. The function $I(X^i_{t}; X^i_{t-\tau})$ is the {\it time-delayed mutual information} (TDMI) with delay $\tau$ for any times series of a stochastic (possibly joint) variable $X_t \in \{X_1, X_2, \ldots, X_T\}$:

\begin{equation}
    I(X_{t}; X_{t-\tau}) = \sum_{i=1}^n p(X_{t}, X_{t-\tau})\log\Big[\frac{ p(X_{t}, X_{t-\tau})}{ p(X_{t}) p(X_{t-\tau})}\Big]. \label{tdmi}
\end{equation}
Equation~\ref{tdmi} encodes the number of bits that variable $X_t$ is able to predict about its own future state based on its ($\tau$-lagged) past states. For example, in a two-agent system, if $X_t = \{X^1_t,X^2_t\}$ then $I(X_{t}; X_{t-\tau})$ is the amount of predictive information data $\tau$ steps in the past is encoded in the current state of the entire joint state of the system. On the other hand, if $X_t = X^1_t$ then $I(X_{t}; X_{t-\tau})$ encodes how much information the past of one agent encodes about its current behaviour. The difference between the whole system TDMI and the sum of the individual's TDMI is the extent to which information is being exchanged between the agents, i.e. $\phi(X;\tau)$ in Equation~\ref{coin} is the {\it excess TDMI}. 

Table~\ref{monk_comp} shows the results of these computations. We note the fact that, as the computer algorithm becomes more sophisticated, from algorithm 0 to 1 to 2, $\phi(X;\tau)$ decreases as the monkey plays closer and closer to the Nash strategy, but then as the sophistication increases, the monkey looks further into the past data to extract information. This is most notable in the difference for algorithm 2: at $\tau = 3$ (0.0148 bits) is nearly twice that of $\tau = 1$ (0.008 bits). This is because, in this specific case, the monkey decouples itself from the computer in order to earn its highest reward: as the computer strategy becomes more sophisticated in measuring the monkeys use of the computer's choices, the less coupled to the computer and its own past the monkey needs to be.

\begin{table}[htbp]
\centering
\begin{tabular}{@{}cccccccccc@{}}
\toprule
Time delay: $\tau$ & Algo & Joint TDMI & Monkey TDMI & Computer TDMI & Excess TDMI: $\phi(X;\tau)$ \\
\midrule
1 & 0 & 0.0999 & 0 & 0 & 0.0999 \\
2 & 0 & 0.1197 & 0.0075 & 0.0002 & 0.1120 \\
3 & 0 & 0.1237 & 0.0095 & 0.0095 & 0.1047 \\
\midrule
1 & 1 & 0.0693 & 0.0011 & 0.0001 & 0.0681 \\
2 & 1 & 0.0748 & 0.0025 & 0.0001 & 0.0722 \\
3 & 1 & 0.0766 & 0.0029 & 0.0005 & 0.0732 \\
\midrule
1 & 2 & 0.0105 & 0.0005 & 0.0002 & 0.0080 \\
2 & 2 & 0.0155 & 0.0019 & 0.0023 & 0.0113 \\
3 & 2 & 0.0216 & 0.0030 & 0.0038 & 0.0148 \\
\bottomrule
\end{tabular}
\caption{The simplest example of two agents interacting with one another via game theory in which $\phi(X;\tau)$ is non-zero. In general we cannot know the details of the flow of information, only that there is a net positive flow across all agents. All values are in bits and computations carried out using JIDT~\cite{lizier2014jidt}. \label{monk_comp}}
\end{table}

\subsection{A triadic example of evolutionary learning at long time scales}

Before we introduce ToM for social interactions, we consider a second example where evolution has found an agent-to-agent interaction that is similar to the worked example used next in Section~\ref{4.0}. Our evolutionary example is a three-agent system: the larval stage of the fly {\it Liriomyza huidobrensis}, the pea plant family {\it Fabaceae} that fly larvae predate on, and the parasitic wasp {\it Opius dissitus} that predates on {\it L. huidobrensis}. These species interact in the following way~\cite{wei2007plants}: The larvae of {\it L. huidobrensis} infest a pea plant, the pea plant gives off volatiles, called {\it infochemicals}, that attract the wasps to the plant, which in turn feed on the larvae, and this larva--wasp conflict indirectly benefits the pea plant. 

In this triadic relationship, the pea plant does not directly respond to the threat from the larvae---for example it has not evolved a chemical agent that repels the larvae-laying flies. Instead it signals a third party, the wasp, to bring the wasp into contact with the larvae, and the wasp then eats the larvae. The wasps and the flies are in a dyadic evolutionary competition that could be modelled using two-agent evolutionary game theory, but the pea plant, having facilitated an instance of this conflict, benefits indirectly from the wasps success, in a sense plants can employ other species as a kind of `body guard'~\cite{kobayashi2007evolution}. We note that, like the example in Section~\ref{4.0}, the plant only signals the wasps to a dyadic interaction when the plant detects the larvae, so that the plant only signals wasps when the plant perceives an intermittent information carrying cue from its environment that wasps are needed.

This type of {\it second order} interaction occurs in other ecological examples as well~\cite{golubski2016ecological} where Trait-Mediated Indirect Interactions (TMIIs) induce hyper-graphs of interactions between species. More complicated interactions have also been observed in which a plant that is being attacked emits infochemicals that lead to {\it unaffected} plants emitting volatiles to attract predators~\cite{kobayashi2007evolution}, to reduce the risk of the unaffected plant being attacked while also aiding the plant being attacked. Kobayashi and Yamamura~\cite{kobayashi2007evolution} specifically call this evolutionary development a form of {\it altruism} but of course there is no cognitive aspect to this altruism. These examples illustrate that evolution produces forms of sophisticated, inter-species, mixed competitive-altruistic interaction networks but without any need for a ToM, individual awareness, or strategic understanding of the interactions, and so individuals are not knowingly strategic or altruistic as we might interpret a person to be. But these evolved strategies are limited, by definition, to be fixed within the lifetime of a single agent and they can only adapt on evolutionary time scales. This is in contrast to a ToM which allows us to adapt to multiple strategic and social contexts that may require novel solutions within the lifespan of a single person. 

n the following section we illustrate how the strategic modification of an interaction network leads to improved CI, contingent on hidden cognitive variables. This provides a formal connection between hidden variables and the causal consequences of network plasticity at the level of the collective. As this is a `causal model' the agent is using, it provides a useful toy model of how an AI may use cognitive tools similar to those of a human to connect the causal use of language with the consequences of network adaptation for improved collective outcomes.

\section{Social network modification through ToM: a minimal model \label{4.0}}

\begin{figure}[h]
    \centering
    \includegraphics[width=0.55\textwidth]{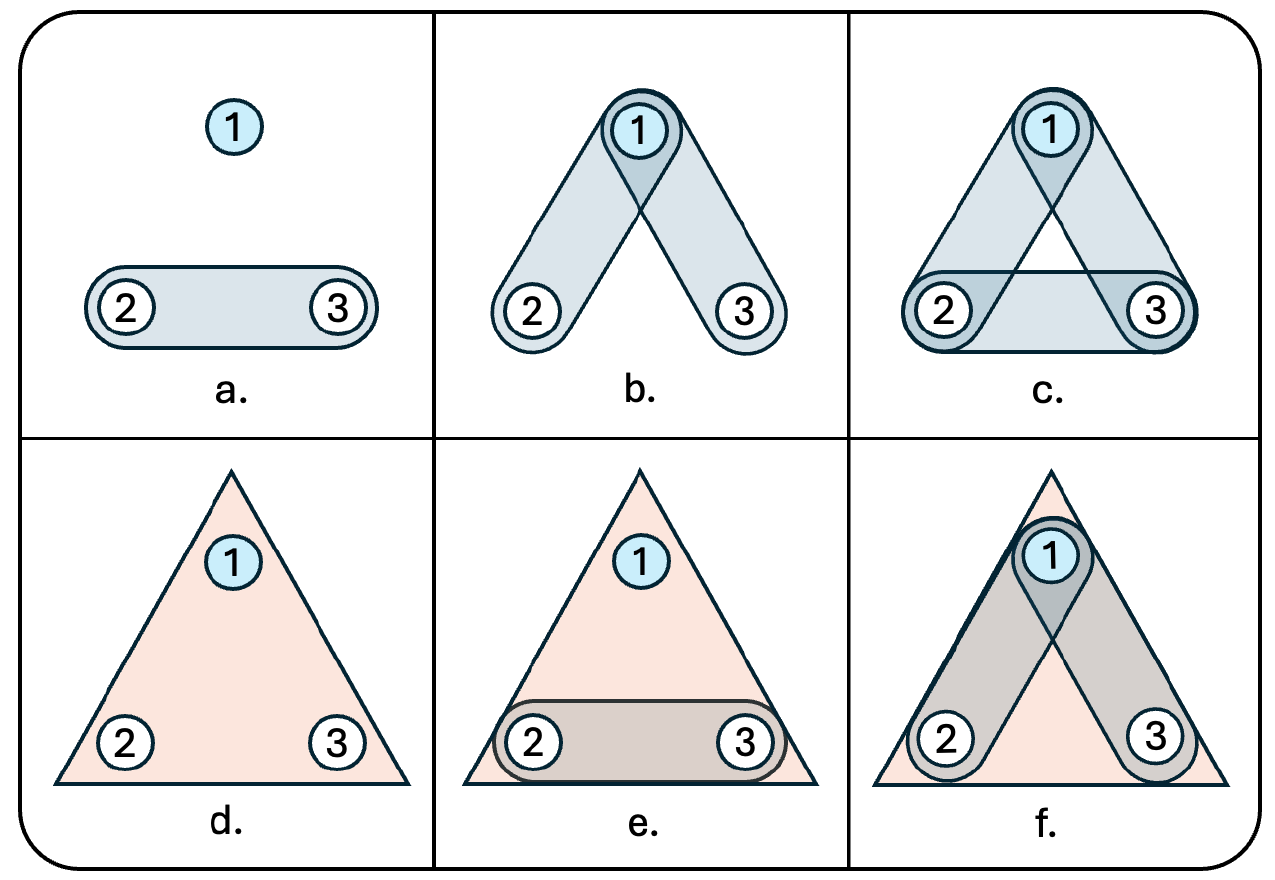}
    \caption{Graphs and hyper-graphs of three agents: (a) A disconnected graph containing two dyadically connected agents and one isolated agent. (b) A connected dyadic graph. (c) A dyadic {\it complete graph}. (d) A hyper-graph in which all three agents are connected by a single hyper-link. (e) A combination of a hyper-graph connecting all agents and a disconnected dyadic graph. (f) A connected dyadic graph and a single hyper-link.} 
    \label{fig:graphs}
\end{figure}

In this model we introduce a simple example of interacting agents that form a hypergraph~\cite{golubski2016ecological}. As in the previous examples, this model is general enough to allow the agents to have any form of biological or artificial psychology, to have zeroth order ToM or fourth order ToM, and it applies to artificial collectives just as readily as it can to biological and ecological collectives.

Game theory provides a formal approach to the analysis of incentivised social interactions in which agents attempt to optimise the outcome (utility) of their joint actions. The expected utilities can be rewritten as polynomials in the agents' decision variables, usually interpreted as probabilistic weights, $x_i \in \mathbf{x}$ for which the co-factors $\mathbf{a}$ are derived from a game payoff matrix (as described in Appendix~\ref{apx:utility-polynomials}). These can be generalised to higher order polynomials representing higher order interactions between the agents: in these expanded utilities the quadratic terms represent the dyadic interactions between agents, the cubic terms represent the three-way interactions, and so on: 
\begin{align}
    U_{i}(\mathbf{x}; \mathbf{a}) &= 
        a^0_i 
        + \sum_j a_i^j x_j 
        + \sum_{j,k} a_i^{jk} x_j x_k 
        + \sum_{j,k,l} a_i^{jkl} x_j x_k x_l
        + \textrm{h.o.t.} \label{eq:1_U_i}
\end{align}
We note that in general $\frac{\partial U_{j}(\mathbf{x}; \mathbf{a})}{\partial x_i}$ describes the impact that $i$'s choice has on agent $j$'s utility and that we need not assume that there exist either symmetrical or even reciprocal impacts between agents' utilities and their behavioural choices. In Figure~\ref{fig:graphs}, we illustrate how three agents are connected via dyadic relationships and higher order (cubic) hyper-graph interactions. Taking the utility for Agent 1 in Figure~\ref{fig:graphs} as an example, the most general description of the interactions in the dyadic utilities for the top row (a)--(c) are given by:
\begin{align}
    (\textrm{a}): \,\, 
        U_1(\mathbf{x}; \mathbf{a}) &= a_1^0 + a_1^1 x_1  &\,\, 
    (\textrm{b}) \, \textrm{and} \, (\textrm{c}): \,\, 
        U_1(\mathbf{x}; \mathbf{a}) &= a_1^0 + \sum_j a_1^j x_j + \sum_{j,k} a_1^{jk} x_j x_k
\end{align}
In (a), Agent 1 is the only agent that can influence their utility, but for (b) and (c), they are also influenced by other agents, linearly and possibly quadratically, depending on the payoff structure. The co-factors $\mathbf{a}$ of these three utilities are derivable directly from the utility matrices of dyadic agent-to-agent interactions in conventional game theory~\cite{harris2023smooth,harre2018multi} (see Appendix~\ref{apx:utility-polynomials}). For utilities (d)--(f) in Figure~\ref{fig:graphs}:
\begin{align}
    (\textrm{d}): \,\, 
        U_1(\mathbf{x}; \mathbf{a}) &= \sum_{j,k,l} a_1^{jkl} x_j x_k x_l 
    &\,\, (\textrm{e}): \,\, 
        U_1(\mathbf{x}; \mathbf{a}) &= a_1^0 + a_1^1 x_1 + \sum_{j,k,l} a_1^{jkl} x_j x_k x_l
\end{align}
\begin{align}
    (\textrm{f}): \,\, U_1(\mathbf{x}; \mathbf{a}) &= 
        a_1^0 
        + a_1^1  x_1 
        + \sum_{j,k} a_1^{jk} x_j x_k 
        + \sum_{j,k,l} a_1^{jkl} x_j x_k x_l
\end{align}
With these descriptions of higher order interactions we can now provide an illustrative example of how a ToM may be used to reconfigure a network to enhance the computational processing of information to improve collective performance.
In the example of the parasitic wasp and the fly larvae eating the leaves of a plant, the plant benefits indirectly from the direct interaction between two other agents. In this case the plant encourages the wasp to come into contact with the larvae in a fight for survival that the plant does not directly participate in. In the example that follows next, a similar but theoretical scenario between three agents is studied where two agents are initially interacting but not producing anything with a third agent on the sidelines. Then the third agent provides a signal to the other two to selectively change their behaviour that directly benefits two of the agents and indirectly benefits the third.

\subsection{Model scenario}

We begin with three agents in proximity to one another situated in a noisy environment in which it is possible for them to collectively do something useful but they are initially in the unfortunate situation in which the behaviour of the three agents produces nothing that is of value. The possible actions ($x_i$) of the agents ($A_i$) are binary: $x_i \in \{-1,1\}$, $i \in \{1,2,3\}$ and so in this example the $x_i$ are not probabilities. $A_1$ is randomly and uniformly changing its states due to a (useful, information carrying) signal it receives from the environment at time $t$: $s^t \in \{-1,1\}$ such that $P(s^t = 1) = P(s^t = -1) = 0.5$. $A_2$ and $A_3$ are initially engaged in the prisoner's dilemma (PD) game where, as a consequence of selfishly (na\"ively) optimising their choice of $x_i$, they are in the defect-defect Nash equilibrium (NE), and so their joint action is constant (with zero value). They are capable of a second output when in the cooperate-cooperate configuration, and this output has a positive value. However, being stuck in the PD NE, they are not initially able to produce it. 

$A_2$ and $A_3$ can produce something of value when they cooperate, but it only has value if the external signal $s^t = +1$. The output value of $A_2$ and $A_3$ interacting is initially a sequence of $0$s: they are not cooperating with each other, and nothing of value is being produced. We assign each agent a state at time $t$, $x^t_i \in \{-1,1\}$, such that their dynamic is an ordered sequence of binary states $[x^1_i, x^2_i, \ldots, x_i^T]$ for $t\in\{1,\ldots, T\}$. The output from $A_2$ and $A_3$ interacting at time $t$ is a result of having matched their states: $V(o^t) = 1$ if $x^t_2 = x^t_3 = 1$ and $V(o^t) = 0$ if $x^t_2 = x^t_3 = -1$, the $x^t_2 \ne x^t_3$ cases are never achieved as they are not NE and $A_2$ and $A_3$ will always choose according to the NE of their interactions. The collective behavioural vector is: $\mathbf{x} = \{x_1, x_2, x_3\}$ or time indexed: $\mathbf{x}^t = \{x_1^t, x_2^t, x_3^t\}$.

\begin{figure}[h]
    \centering
    \includegraphics[width=0.8\textwidth]{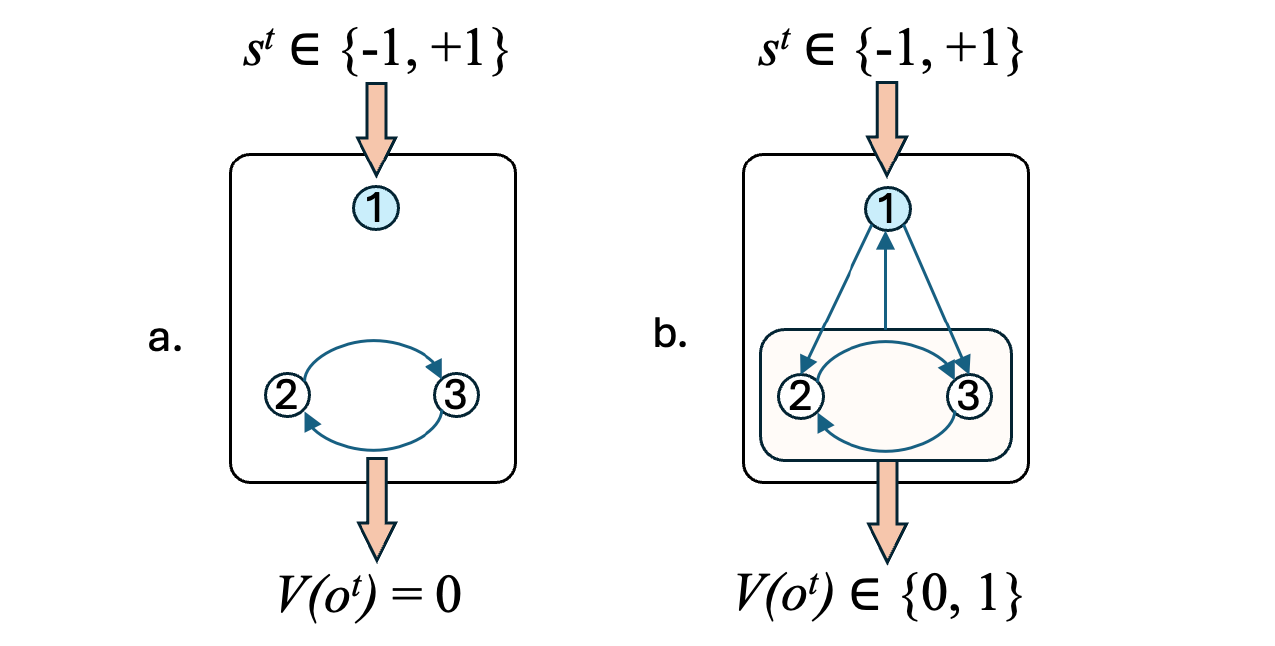}
    \caption{The interaction networks for the two scenarios: 
        (a) $A_1$ perceives signal $s^t$ but cannot relay it to $A_2$ and $A_3$ and these two agents are not cooperating and so produce $0$ value from their output $o^t$. 
        (b) $A_1$ influences $A_2$ and $A_3$ to cooperate when $A_1$ receives a signal from the environment that cooperating will result in a positive payoff for $A_2$ and $A_3$; as a consequence, $A_1$ receives a portion of the utility from both $A_2$ and $A_3$.} 
    \label{fig:example}
\end{figure}

The signal $s^t$ that $A_1$ receives indicates whether or not $A_2$ and $A_3$  should cooperate at time $t$, but initially no information is passing from $A_1$ to $\{A_2, A_3\}$ and $A_1$ cannot produce anything by itself. So $U_1(\mathbf{x};\mathbf{a}) = 0$, and agents $A_2$ and $A_3$ only produce output of zero value: $V(o^t) = 0$. The agent utilities in this case are:
\begin{align}
U_1(\mathbf{x};\mathbf{a}) &= 0, &\,\,
U_2(\mathbf{x};\mathbf{a}) &= a_2^0 + a_2^3 x_3 + a_2^2 x_2, &\,\,
U_3(\mathbf{x};\mathbf{a}) &= a_3^0 + a_3^2 x_2 + a_3^3 x_3,
\end{align}
and $U_2 = U_3 = 0$ for defect-defect joint strategies. This is equivalent to the configuration in Figure~\ref{fig:graphs}(a), and we illustrate this in Figure~\ref{fig:example}, a. In the second scenario, we assume that $A_2$ and $A_3$ are still na\"ively pursuing their selfish goals, but now $A_1$ is more psychologically aware: it knows the beliefs, preferences, and constraints of the other two agents, and it is able to influence their preferences so that they will cooperate with each other when $A_1$ receives the signal: $s^t = 1$.
They, in turn, will pay $A_1$ some portion of their total payoff.

\begin{figure}[tp]     
    \centering
    \begin{tabular}{cc}
            \begin{subtable}[t]{0.4\textwidth}
                \centering
                \begin{tabular}{|c|l|c|c|}
                    \cline{3-4}
                    \multicolumn{2}{c|}{} & \multicolumn{2}{c|}{$A_3 \rightarrow x_3$} \\ \cline{3-4}
                    \multicolumn{2}{c|}{} & $+1 \equiv \texttt{C}$ & $-1\equiv \texttt{D}$ \\ \hline
                    \multirow{2}{*}{$A_2 \rightarrow x_2$} & $+1\equiv \texttt{Cooperate}$ & (R, R) & (S, T) \\ \cline{2-4}
                    & $-1\equiv \texttt{Defect}$ & (T, S) & (P, P) \\ \hline
                \end{tabular}
                \caption{Generalised payoff matrix}
            \end{subtable} 
         &
        \begin{subtable}[t]{0.55\textwidth}
            \centering
            \begin{tabular}{|l|c|c|}
                \cline{2-3}
                \multicolumn{1}{c|}{} & \texttt{C} & \texttt{D} \\ \hline
                \texttt{C} & $(1,1)$ & $(0-c,1+c)$ \\ \hline
                \texttt{D} & $(1+c,0-c)$ & $(0,0)$ \\ \hline
            \end{tabular}
            \caption{Prisoner's Dilemma ($c=-\frac{1}{4}$) and Harmony ($c= \frac{1}{4}$)}
        \end{subtable}
    \end{tabular} 
        \caption{
        Payoff matrices for agents $A_2$ and $A_3$, which agent $A_1$ can strategically influence by setting $c \leftarrow x_1$.
    }
    \label{fig:game-matrices}
\end{figure}

To do this numerically, we standardise the symmetric, two-player, two-choice, game theory payoff matrices for $A_2$ and $A_3$ by setting $R=1$ and $P=0$ (Figure~\ref{fig:game-matrices}, a. and see~\cite{harre2018multi}). We note that the Harmony game (\texttt{Ha}), in which both agents coordinating is the only Nash equilibrium and $T>R>P>S$, can be transformed into the Prisoner's Dilemma (\texttt{PD}), in which both agents defecting is the only Nash equilibrium and $R>T>S>P$, by introducing a two-state parameter $c~\in~\{-\frac{1}{4}, \frac{1}{4}\}~\implies~\{\texttt{PD},\texttt{Ha}\}$ and replacing $T = R + c$ and $S = P -c$ so that the payoff matrices derivable from Figure~\ref{fig:game-matrices},b. result in the utilities:
\begin{align}
    U_2(\mathbf{x};\mathbf{a}) &= R + cx_2 - (R+c)x_3, &\,\,  
    U_3(\mathbf{x};\mathbf{a}) &= R + cx_3 - (R+c)x_2. \label{indie_utils}
\end{align}

As agent $A_1$ is aware of the specific BPC model by which $A_2$ and $A_3$ make their decisions, $A_1$ adjusts their behaviour by manipulating their utility co-factors following $s^t$, by setting $x^t_1 = \frac{1}{4}s^t$ such that $x^t_1 \in \{-\frac{1}{4}, \frac{1}{4}\}$. $A_1$ then modulates the utility of the other two agents (via an information-carrying signal such as speaking to them) and, in response, receives 10\% of the payoffs that result from the interaction between $A_2$ and $A_3$.
The resulting payoffs for each agent are (derived in detail in Appendix~\ref{apx:utility-derivation}):
\begin{align}
    U_1(\mathbf{x};\mathbf{a}) &= \frac{1}{10}\big(U_2(\mathbf{x};\mathbf{a}) + U_3(\mathbf{x};\mathbf{a})\big), &\,\,
    U_2(\mathbf{x};\mathbf{a}) &= R + x_1x_2 - (R+x_1)x_3, &\,\, 
    U_3(\mathbf{x};\mathbf{a}) &= R + x_1x_3 - (R+x_1)x_2. \label{hyper_utils}
\end{align}

\subsection{Model interpretation}

We make three observations regarding these two scenarios. First, we note that the interactions in the second scenario are a hypergraph in the sense that if any one of the nodes were removed in Figure~\ref{fig:example}, b. no remaining agent receives any payoff, all agents are necessary contributors to any single agent receiving a utility for their actions~\cite{golubski2016ecological}. Second, it is straightforward to compute the collective intelligence of both scenarios using Equation~\ref{coin} and the three stochastic variables $\{x_1^t,x_2^t,x_3^t\}$ when $s^t \in \{-1,1\}$ is uniformly distributed: In the first scenario it is 0 bits and for the second scenario it is 1 bit. This is a simple illustration of how a single agent, knowing the BPC model of two other agents, who are both na\"ive to all other agents' BPC models, can improve the collective intelligence of the group by manipulating the network of interactions between agents. This is analogous to the results discussed in Woolley~\cite{woolley2010evidence} in which group performance, using a measure of collective intelligence, is improved by having agents who have a theory of mind. Third, unlike evolutionary processes, the process in the second scenario has a goal-directed agent who is psychologically informed about the internal states of the other agents and so is able to make plans contingent on different configurations of the interaction network. This implies that $A_1$ understands the preferences implied by the co-factors of Equation~\ref{indie_utils}, ($R+c$) and Equation~\ref{hyper_utils}, ($R+x_1$), influencing the behaviour of the other two agents. These preferences are the hidden (cognitive) variables discussed in Section~\ref{2.2}, i.e. the {\it Preferences} aspect of the BPC framework. In terms of levels of ToM, $A_1$ only needs to ascribe hidden states to the other two agents but not the ability for $A_2$ or $A_3$ to ascribe hidden states to $A_1$, for example the NE can be reached by $A_2$ and $A_3$ by observing the other agent's behaviour. So the ToM of $A_1$ is of second order and $A_2$ and $A_3$ only need a first order ToM. In this sense a ToM allows $A_1$ to reconfigure their social networks to further their goals at much shorter time scales than the evolutionary time scale of ecological networks.

\section{Strategic Inference, Theory of Mind, and the Dual Role of \texttt{piKL}}

In strategic multi-agent environments, coordination and collaboration often depend not only on optimal decision-making, but on the agent's ability to reason about the intentions, beliefs, and goals of others. This capability—often referred to as \emph{Theory of Mind} (ToM)—becomes even more powerful when paired with language as a medium for influencing and inferring mental states. In this section, we develop a formal model that integrates reinforcement learning, ToM, and communication into a unified agent architecture. We also clarify two distinct interpretations of the \texttt{piKL} algorithm introduced in the Cicero AI framework.

\subsection{Baseline Reinforcement Learning Policy}

Let $s \in \mathcal{S}$ denote the environment state and $a \in \mathcal{A}$ the action selected by agent $i$. The baseline policy derived from reinforcement learning (RL) maximizes expected return:
\[
\pi^{\text{RL}}(a \mid s) = \arg\max_a Q(s, a)
\]
where $Q(s, a)$ is the value function learned via standard temporal-difference or policy-gradient methods.

\subsection{Theory of Mind and Noisy Communication Channels}

Each agent $i$ is assumed to have an internal, latent cognitive state $\theta_i \in \Theta$, encoding beliefs, goals, or preferences. Suppose agent $i$ communicates with agent $j$ via a message $m \in \mathcal{M}$ drawn from a noisy channel:
\[
P(m \mid \theta_i)
\]
The receiving agent $j$ updates its belief over $\theta_i$ using Bayes' rule:
\[
b_j(\theta_i \mid m) = \frac{P(m \mid \theta_i) \cdot P(\theta_i)}{\sum_{\theta'_i} P(m \mid \theta'_i) \cdot P(\theta'_i)}
\]
This belief over the sender’s internal state allows agent $j$ to modulate its action policy accordingly.

\subsection{ToM-Enhanced Policy via Belief Inference}

The ToM-enhanced policy for agent $j$ is defined by taking the expectation over the inferred mental states of agent $i$:
\[
\pi_j^{\text{ToM}}(a \mid s, m) = \mathbb{E}_{\theta_i \sim b_j(\theta_i \mid m)} \left[ \pi_j(a \mid s, \theta_i) \right]
\]
This formulation permits the agent to adaptively align, cooperate, or compete based on its belief about the intentions or preferences of the other agent.

\subsection{Strategic Pragmatics and Message Selection}

Recognizing that the receiver is performing this inference, the speaker agent $i$ may choose its message strategically to manipulate beliefs:
\[
m^* = \arg\max_m \mathbb{E}_{\theta_j \sim b_i(\theta_j)} \left[ U_i(\theta_j(m)) \right]
\]
This recursive, belief-aware messaging strategy is analogous to pragmatic language models such as the Rational Speech Acts (RSA) framework.

\subsection{Two Interpretations of \texttt{piKL}}

The \texttt{piKL} method measures divergence between policies. We describe two distinct formulations:

\paragraph{1. Anchor-Regularised \texttt{piKL} (Cicero)}

In the Cicero system, the agent's current policy $\pi_i$ is constrained by a divergence from an anchor policy $\pi_{\text{anchor}}$, learned from human gameplay:
\[
D_{\text{piKL}}^{\text{anchor}} = \mathrm{KL}\left[ \pi_i(a \mid s) \parallel \pi_{\text{anchor}}(a \mid s) \right]
\]
This encourages the agent to remain near human-like, interpretable behavior. The total objective becomes:
\[
\mathcal{L}_{\text{anchor}} = \mathbb{E}_{(s,a) \sim \pi_i} \left[ R(s,a) \right] - \lambda_1 \cdot D_{\text{piKL}}^{\text{anchor}}
\]

\paragraph{2. ToM-Induced \texttt{piKL}}

Alternatively, we define a divergence between the baseline RL policy and the ToM-enhanced policy:
\[
D_{\text{piKL}}^{\text{ToM}} = \mathrm{KL}\left[ \pi^{\text{RL}}(a \mid s) \parallel \pi^{\text{ToM}}(a \mid s, m) \right]
\]
This measures how much the agent’s behavior is changed by incorporating belief inference and communicative reasoning.

\subsection{Unified Objective}

We combine both interpretations in a unified loss:
\[
\mathcal{L}_{\text{total}} = \mathbb{E}_{(s,a) \sim \pi_i} \left[ R(s,a) \right]
- \lambda_1 \cdot \mathrm{KL}\left[ \pi_i \parallel \pi_{\text{anchor}} \right]
- \lambda_2 \cdot \mathrm{KL}\left[ \pi^{\text{RL}} \parallel \pi^{\text{ToM}} \right]
\]
Here, $\lambda_1$ and $\lambda_2$ control the influence of human-alignment and social-inference regularisation, respectively. Together, they allow an agent to balance exploitation, interpretability, and strategic communicative inference.

\subsection{Interpretation}

This model formalizes a cognitive architecture in which:
\begin{itemize}
  \item Reinforcement learning drives raw goal-seeking behavior;
  \item Theory of Mind allows adaptation to others' latent beliefs;
  \item Language serves as a noisy yet informative conduit of cognitive intent;
  \item Pragmatic reasoning aligns communicative goals with strategic planning;
  \item Dual regularisation (via \texttt{piKL}) ensures the agent balances optimality with trustworthiness and social intelligence.
\end{itemize}

\subsection{Our perspective in context}
We have reviewed some of the recent advances in how people, as complex agents, with a vast space of specialised, context-dependent behaviours, are able to coordinate their activities. In particular, finding ways to recombine our individual competencies to produce, in a short period of time, the appropriate collective competencies is a combinatorically complex task. The central theme of this article is that having a causal (generative \cite{friston2023free,ruiz2024factorised}) model of another person’s behaviour that reflects the current environmental and social context helps us manage this complexity. This is one way in which our solid brains produce the necessary liquid social structures that quickly build collective solutions with novel collective competencies. In this section we bring these ideas together and summarise our view.

We first draw attention to how, in Section~\ref{4.0}, $A_1$ has constructed a niche for itself in the context of the preexisting dyadic network between $A_2$ and $A_3$. In ecological networks a new agent joins a pre-existing network if it can find a niche within which it can fit. This occurs in one of three distinct ways: niche {\it choice}, niche {\it conformance}, and niche {\it construction}~\cite{clark2020niche,trappes2022individualized}. Niche choice occurs when an individual selects environmental conditions that align with its phenotype, while niche conformance involves adjusting its phenotype to suit the environment. Niche construction is the modification of the environment to meet individual needs, which may also impact other species. This suggests an analogy between the formation of ecological networks, where new agents joining may enhance or disrupt the current configuration, and social networks where membership can be explicitly or implicitly gated according to a prospective agent's `fit' within the group, and even if the group can adjust to accommodate a newcomer, just as the newcomer can adjust in order to be accepted. In human social groups people can be recruited or excluded depending on their contribution to the better functioning of the collective, which in turn corresponds to changes in individual neural activity related to the social network structure~\cite{schmalzle2017brain}.

In our analogy with ecological networks, we suggest that {\it any} signalling agents in a network need to encode messages over an information-carrying medium that receivers are receptive to and that can then be decoded by a specific receiver. We have argued that humans, as both signallers and receivers, take advantage of our ToM in order to understand what signals will be correctly interpreted by another person and to adapt their signalling to the cognitive state of the receiver. Specifically, our psychology allows us to learn how to read, interpret, and re-write our interpersonal communication over a very short time frame, allowing us rapid, targeted control over our collective behaviour.

In some sense the adage {\it there is nothing new under the sun} holds here as evolution and biology have recycled fundamental, pre-existing principles in the service of human sociality. But the adaptive speed of our social networks, the psychological mechanisms involved, the variety of purposes they serve, and the complexity of the communications appear most highly developed in humans. In ecological networks, for example, some agents can act as an encoder-sender of semantic information via infochemicals that signal another agent. A second agent then acts as a receiver-decoder of this signal that in turn changes the behaviour of the receptive agent. The combination of the receiver's anatomical configuration and behavioural phenotype elicits an appropriate stimulus-response that benefits the signalling agent, and these interactions can form vast, complex hypergraphs of competition and cooperation between agents of many different species. But neither agent needs higher cognitive faculties---their ToM is of the zeroth order. By analogy, humans can target specific individuals or groups of individuals in order to have them adjust their behaviours to best suit the goals of the signaller. The competencies that a ToM affords the sender allows them to know that a receiver is capable of both decoding the signal and acting appropriately in response because they know of the receiver's {\it receptive} and {\it causal} states, both cognitive (hidden) and behavioural (overt). That is to say they take advantage of their ToM to understand how, when, what, and to whom they need to signal in order to achieve a beneficial outcome, whereas blind evolution would take much longer to achieve the same results.


The separation of time scales also plays an important distinction between learning processes that use the same theoretical foundation. The mathematical description of the processes that underpin the evolution of species has been shown to be equivalent to the statistical learning process that underpins Bayesian learning in individual agents~\cite{campbell2016universal,watson2016evolutionary,watson2016can}, for both biological and artificial agents. What Bayesian learning provides at the level of the individual agent is an advantage in the speed of adaptation that Bayesian learning via evolution cannot achieve. Again, this suggests a universality of description that only varies qualitatively in terms of time scales and the specifics of the mechanisms.

The models we presented highlight the roles ToM plays in human social collectives such as deciding which relationships to develop and how these relationships aid in the collective processing of information towards an end goal. Being able to understand, predict, and manipulate these information-carrying connections is valuable in improving group dynamics (in human groups, AI groups, or hybrid human-AI groups) and for the future design of ToM-based AI. 
For example, knowing that ToM plays a role in group performance \cite{woolley2010evidence}, quantifying and modelling (ToM-mediated) CI will shed light on the types of interventions or adjustments that can modulate it. Additionally, including AI agents with sophisticated ToM in human groups has the potential to `construct social niches' (see below) more effectively, such as by modifying the incentives involved in a social setting (making some actions more desirable than others; cf. mechanism design~\cite{phelps2010evolutionary}), or the connections between agents~\cite{McKee2023ScaffoldingCooperation}, or by simply mediating interactions with an increased capacity for memory, attention, and reasoning \cite{Westby2023CollectiveIntelligence}.
Recent efforts have leveraged ToM and AI for improving team effectiveness \cite{Bendell2024IndividualTeam, Westby2023CollectiveIntelligence} and understanding the link between ToM and deception \cite{Sarkadi2023ArmsRace}, or ToM and expectation formation in markets \cite{Bao2024ReadingMarket}.
Others suggest AI with ToM could assist humans in improving their ToM and communications skills \cite{Garcia-Lopez2023TheoryMind}, in negotiation, education (adaptation to students needs and knowledge), and games \cite{Rocha2023ApplyingTheory}; healthcare (to tailor care to patients' mental states), self-driving cars (anticipating the actions of other cars), workplaces (employee mood and stress), and marketing \cite{Nebreda2023SocialMachine}.

To successfully apply ToM-based AI to these areas, we need to consider how we could define the appropriate incentives for an AI agent to adapt to a given set of social circumstances in a desirable manner. Our examples, while minimal, offer a conceptual framing for thinking about this human-AI integration. For example, the use of information theory to quantify the information processing provides a measure that can be used in an optimisation objective for an AI agent within a collective, incentivising an AI to act in ways that are constructive for the collective as a whole. 

The difficulties of implementing socially aware algorithms in an AI are also made more complicated by human variations in expectations. For example, the famous trolley problem has often been used as an example of a moral dilemma that an autonomous AI that is in control of a car may confront during an emergency. To understand the complexities of this issue for an AI, Awad et al~\cite{awad2018moral} collected ~40 million individual decisions from 233 countries and territories in 10 languages. The questions focused on trolley-like moral dilemmas to examine cross-cultural ethical variation, showing three major clusters of countries and that these variations reflect differences in modern institutions and deep cultural traits. These results illustrate that our morality is not universal, there is a rich variety of expectations for the same moral dilemmas, and that any AI would need to reflect the local cultural variations in which they are used. In the Discussion next we cover the broader issue of how an AI will need to learn and adapt in order to fit the social circumstances the AI will be expected to be constrained by, very much like other socio-ecological systems.

To date the study of the collective intelligence in hybrid AI-human systems is in its infancy, however we already have many of the necessary tools to begin these investigations. ToM is experimentally well established in psychology and work has begun developing and implementing algorithms for an AI-ToM, see for example Mao et al's recent review of empirical work on beliefs, desires and preferences for AI with ToM~\cite{mao2024review} as well as the computational, algorithmic, and experimental articles discussed in the Introduction above. What is not as well understood are the dynamics that agential AI will introduce into our human social ecology, and to that extent the study of {\it machine behaviour}~\cite{rahwan2019machine} and new measures of collective intelligence as we have described here need to be applied to AI-human hybrid systems to better understand the collective outcomes, rather than focusing on the algorithmic foundations.

\section{Discussion: Theory of Mind in the Context of Social AI \label{5.0}}

Each step in our cultural development has changed the ways in which we combine individual skills to achieve better, more sophisticated collective outcomes. There is evidence that hunter gatherers participated in the division of labour, complex social networks, multilevel, and fractal-like social structures long before we settled into villages~\cite{de2022understanding,dyble2016networks,harre2016social,hamilton2007complex}. This was in part due to our ability to construct our own niches~\cite{laland2016introduction} and this practice has persisted from hunter-gathers to farming~\cite{rowley2011foraging} through to civilisation building. As Arroyo-Kalin {\it et al.}~\cite{arroyo2017civilisation} quote in their opening to the special issue {\it Civilisation and Human Niche Construction}:
\begin{quote}
It is impossible to avoid the conclusion that organisms construct every aspect of their environment themselves. They are not the passive objects of external forces, but the creators and modulators of these forces. The metaphor of adaptation must therefore be replaced by one of construction, a metaphor that has implications for the form of evolutionary theory (Levins and Lewontin 1985: 104).
\end{quote}
It has also been argued that through the manipulation of our environmental niche we have brought about the Anthropocene~\cite{kemp20207000,smith2013onset,ellis2024anthropocene,wilson2023multilevel}.

On the other hand, {\it social niche construction}~\cite{ryan2016social} extends this idea to the process whereby agents modify their social context so as to influence their own social evolution. A {\it social niche} is the context in which social behaviour occurs, and so social niche construction is where agents actively choose to change their social environments, for example, by choosing who to associate with and how to behave~\cite{laland2016introduction}. Ryan {\it et al.}~\cite{ryan2016social} describe this in game theoretical terms as the {\it effective game} agents are playing after all relevant factors are accounted for, such as the underlying game itself (the payoff matrices in conventional games for example) and any {\it social niche modifiers}, amongst others. A {\it social niche modifier} is a trait that alters the effective game being played, causing it to differ from the immediate payoff matrices that are usually the complete description of the incentivised interactions, for example population structure, relatedness, punishment etc. In the example in Section~\ref{4.0}, $A_1$ constructs a social niche for itself by manipulating the structure of the game that $A_2$ and $A_3$ are playing, for the benefit of $A_1$ and incidentally for the benefit of the other agents as well. However, social niche construction theory pertains to evolution more broadly and is not specific to human social networks as the formal description of the model in Section~\ref{4.0} could be an evolved network of plants and animals or a more rapidly changing social network.

In moving towards AI that is situated within a hybrid human-AI ecology, the complexities of effective network construction, communication, and cognitive tools will need to be worked through. Here, we discuss just two of the issues: the difficulties of building psychologically complex AIs and the social environment AI will need to adapt to. The capacity of current AI theories to be sufficient to encompass, in principle at least, human levels of cognition is a rich area of research, from attention mechanisms~\cite{vaswani2017attention}, to reinforcement learning subsuming reasoning~\cite{silver2021reward} and chain of thought reasoning in language models~\cite{li2024chain} there are arguments being made for {\it emergent phenomena} in AI~\cite{wei2022emergent} as well as ToM~\cite{strachan2024testing} and even artificial general intelligence~\cite{bubeck2023sparks}. What is often missed in these examples is that the individual AI's ability to digest information and update parameter weights is only a small fraction of what is needed to be effective in a specifically social context. An important tool in our psychological toolbox is our ability to maintain shared hidden variables that are the causal basis of our coordinated joint actions in physical and social environments. Even progress on single agents having effective causal models of the physical environment has been much slower than in other areas of AI~\cite{xiong2022artificial}. As we have shown in this article, there is very good evidence that causal models, both physical and social, are necessary for people to be able to communicate with each other, and communication is inextricably tied to our ToM. This triad of language, shared causal models of hidden variables, and ToM appears to be a minimum for human social coordination. 

Even with this triad established within an AI, the next challenge is where, when, and how an AI should fit into any given collaborative social context. Simple machine intelligence in collaboration with people, such as auto-correct, recommender systems, or GPS navigation, are effective because a human decided there was a need for these tools, they placed them in the appropriate context and then the users adjust their behaviour to any shortcomings the tool might have. But the more autonomous machine intelligence becomes, the more complex the physical and social environment is, and the more trust that needs to be placed in the causal models that drive the behaviour of an AI (i.e. their analogue of a person's hidden cognitive variables) the more an AI needs to know how to interact with us in a way that resembles how we interact with each other. They will ultimately need to be able to do this via niche construction, niche adaptation, and niche choice, all of which, for people, is a negotiated relationship between each other that needs to be satisficing for those involved. This brings human-AI joint adaptation closer to what Laland {\it et al.} have in mind when considering a rethink of evolutionary theory~\cite{laland2014does}:
\begin{quote}
We hold that organisms are constructed in development, not simply ‘programmed’ to develop by genes. Living things do not evolve to fit into pre-existing environments, but co-construct and coevolve with their environments, in the process changing the structure of ecosystems.
\end{quote}

Consequently, we argue that for any AI to be a {\it beneficially adaptive}, autonomous, and socially aware agent, it will also need to reflect the universal principles we see in many evolutionary transitions~\cite{szathmary1995major,szathmary2015toward,prokopenko2024biological}. This includes multi-level selection in economic~\cite{wilson2023multilevel,wilson2024rethinking} and biological systems~\cite{czegel2019multilevel,levin2023darwin} as well as the hierarchical transitions~\cite{czegel2019multilevel} that appear to be ubiquitous in our biological, social, economic, and technological history. 

\bibliographystyle{unsrt}
\bibliography{references}

\begin{thebibliography}{100}

\bibitem{levin2023bioelectric}
Michael Levin.
\newblock Bioelectric networks: the cognitive glue enabling evolutionary scaling from physiology to mind.
\newblock {\em Animal Cognition}, 26(6):1865--1891, 2023.

\bibitem{momennejad2022collective}
Ida Momennejad.
\newblock Collective minds: social network topology shapes collective cognition.
\newblock {\em Philosophical Transactions of the Royal Society B}, 377(1843):20200315, 2022.

\bibitem{migliano2022origins}
Andrea~Bamberg Migliano and Lucio Vinicius.
\newblock The origins of human cumulative culture: from the foraging niche to collective intelligence.
\newblock {\em Philosophical Transactions of the Royal Society B}, 377(1843):20200317, 2022.

\bibitem{hamilton2007complex}
Marcus~J Hamilton, Bruce~T Milne, Robert~S Walker, Oskar Burger, and James~H Brown.
\newblock The complex structure of hunter--gatherer social networks.
\newblock {\em Proceedings of the Royal Society B: Biological Sciences}, 274(1622):2195--2203, 2007.

\bibitem{hill2008network}
Russell~A Hill, R~Alexander Bentley, and Robin~IM Dunbar.
\newblock Network scaling reveals consistent fractal pattern in hierarchical mammalian societies.
\newblock {\em Biology letters}, 4(6):748--751, 2008.

\bibitem{harre2016social}
Michael~S Harr{\'e} and Mikhail Prokopenko.
\newblock The social brain: scale-invariant layering of erd{\H{o}}s--r{\'e}nyi networks in small-scale human societies.
\newblock {\em Journal of the Royal Society Interface}, 13(118):20160044, 2016.

\bibitem{sole2019liquid}
Ricard Sol{\'e}, Melanie Moses, and Stephanie Forrest.
\newblock Liquid brains, solid brains, 2019.

\bibitem{pinero2019statistical}
Jordi Pi{\~n}ero and Ricard Sol{\'e}.
\newblock Statistical physics of liquid brains.
\newblock {\em Philosophical Transactions of the Royal Society B}, 374(1774):20180376, 2019.

\bibitem{levin2019computational}
Michael Levin.
\newblock The computational boundary of a “self”: developmental bioelectricity drives multicellularity and scale-free cognition.
\newblock {\em Frontiers in psychology}, 10:2688, 2019.

\bibitem{levin2022technological}
Michael Levin.
\newblock Technological approach to mind everywhere: an experimentally-grounded framework for understanding diverse bodies and minds.
\newblock {\em Frontiers in systems neuroscience}, 16:768201, 2022.

\bibitem{levin2023darwin}
Michael Levin.
\newblock Darwin’s agential materials: evolutionary implications of multiscale competency in developmental biology.
\newblock {\em Cellular and Molecular Life Sciences}, 80(6):142, 2023.

\bibitem{fields2022competency}
Chris Fields and Michael Levin.
\newblock Competency in navigating arbitrary spaces as an invariant for analyzing cognition in diverse embodiments.
\newblock {\em Entropy}, 24(6):819, 2022.

\bibitem{hopfield1982neural}
John~J Hopfield.
\newblock Neural networks and physical systems with emergent collective computational abilities.
\newblock {\em Proceedings of the national academy of sciences}, 79(8):2554--2558, 1982.

\bibitem{beniaguev2021single}
David Beniaguev, Idan Segev, and Michael London.
\newblock Single cortical neurons as deep artificial neural networks.
\newblock {\em Neuron}, 109(17):2727--2739, 2021.

\bibitem{deneve2008bayesian}
Sophie Deneve.
\newblock Bayesian spiking neurons i: inference.
\newblock {\em Neural computation}, 20(1):91--117, 2008.

\bibitem{kay2024ant}
Tomas Kay, Alba Motes-Rodrigo, Arthur Royston, Thomas~O Richardson, Nathalie Stroeymeyt, and Laurent Keller.
\newblock Ant social network structure is highly conserved across species.
\newblock {\em Proceedings B}, 291(2027):20240898, 2024.

\bibitem{richardson2021leadership}
Thomas~O Richardson, Andrea Coti, Nathalie Stroeymeyt, and Laurent Keller.
\newblock Leadership--not followership--determines performance in ant teams.
\newblock {\em Communications biology}, 4(1):535, 2021.

\bibitem{stockmaier2021infectious}
Sebastian Stockmaier, Nathalie Stroeymeyt, Eric~C Shattuck, Dana~M Hawley, Lauren~Ancel Meyers, and Daniel~I Bolnick.
\newblock Infectious diseases and social distancing in nature.
\newblock {\em Science}, 371(6533):eabc8881, 2021.

\bibitem{galesic2023beyond}
Mirta Galesic, Daniel Barkoczi, Andrew~M Berdahl, Dora Biro, Giuseppe Carbone, Ilaria Giannoccaro, Robert~L Goldstone, Cleotilde Gonzalez, Anne Kandler, Albert~B Kao, et~al.
\newblock Beyond collective intelligence: Collective adaptation.
\newblock {\em Journal of the Royal Society interface}, 20(200):20220736, 2023.

\bibitem{mirowski1998markets}
Philip Mirowski and Koye Somefun.
\newblock Markets as evolving computational entities.
\newblock {\em Journal of Evolutionary Economics}, 8(4):329--356, 1998.

\bibitem{l2003economics}
Robert L.~Axtell.
\newblock Economics as distributed computation.
\newblock In {\em Meeting the Challenge of Social Problems via Agent-Based Simulation: Post-Proceedings of the Second International Workshop on Agent-Based Approaches in Economic and Social Complex Systems}, pages 3--23. Springer, 2003.

\bibitem{mirowski2007markets}
Philip Mirowski.
\newblock Markets come to bits: Evolution, computation and markomata in economic science.
\newblock {\em Journal of Economic Behavior \& Organization}, 63(2):209--242, 2007.

\bibitem{harre2022entropy}
Michael~S Harr{\'e}.
\newblock Entropy, economics, and criticality.
\newblock {\em Entropy}, 24(2):210, 2022.

\bibitem{harre2009phase}
Michael~S. Harr{\'e} and Terrence Bossomaier.
\newblock Phase-transition--like behaviour of information measures in financial markets.
\newblock {\em Europhysics Letters}, 87(1):18009, 2009.

\bibitem{harre2015entropy}
Michael Harr{\'e}.
\newblock Entropy and transfer entropy: the dow jones and the build up to the 1997 asian crisis.
\newblock In {\em Proceedings of the International Conference on Social Modeling and Simulation, plus Econophysics Colloquium 2014}, pages 15--25. Springer International Publishing, 2015.

\bibitem{langton1990computation}
Chris~G Langton.
\newblock Computation at the edge of chaos: Phase transitions and emergent computation.
\newblock {\em Physica D: nonlinear phenomena}, 42(1-3):12--37, 1990.

\bibitem{moore2018inform}
Douglas~G Moore, Gabriele Valentini, Sara~I Walker, and Michael Levin.
\newblock Inform: efficient information-theoretic analysis of collective behaviors.
\newblock {\em Frontiers in Robotics and AI}, 5:60, 2018.

\bibitem{wibral2014directed}
Michael Wibral, Raul Vicente, and Joseph~T Lizier.
\newblock {\em Directed information measures in neuroscience}, volume 724.
\newblock Springer, 2014.

\bibitem{tononi2016integrated}
Giulio Tononi, Melanie Boly, Marcello Massimini, and Christof Koch.
\newblock Integrated information theory: from consciousness to its physical substrate.
\newblock {\em Nature reviews neuroscience}, 17(7):450--461, 2016.

\bibitem{mediano2022integrated}
Pedro~AM Mediano, Fernando~E Rosas, Juan~Carlos Farah, Murray Shanahan, Daniel Bor, and Adam~B Barrett.
\newblock Integrated information as a common signature of dynamical and information-processing complexity.
\newblock {\em Chaos: An Interdisciplinary Journal of Nonlinear Science}, 32(1), 2022.

\bibitem{barrett2011practical}
Adam~B Barrett and Anil~K Seth.
\newblock Practical measures of integrated information for time-series data.
\newblock {\em PLoS computational biology}, 7(1):e1001052, 2011.

\bibitem{mediano2018measuring}
Pedro~AM Mediano, Anil~K Seth, and Adam~B Barrett.
\newblock Measuring integrated information: Comparison of candidate measures in theory and simulation.
\newblock {\em Entropy}, 21(1):17, 2018.

\bibitem{prokopenko2009information}
Mikhail Prokopenko, Fabio Boschetti, and Alex~J Ryan.
\newblock An information-theoretic primer on complexity, self-organization, and emergence.
\newblock {\em Complexity}, 15(1):11--28, 2009.

\bibitem{lizier2012local}
Joseph~T Lizier.
\newblock {\em The local information dynamics of distributed computation in complex systems}.
\newblock Springer Science \& Business Media, 2012.

\bibitem{bossomaier2016transfer}
Terry Bossomaier, Lionel Barnett, Michael Harr{\'e}, and Joseph~T Lizier.
\newblock {\em An Introduction to Transfer Entropy: Information Flow in Complex Systems}.
\newblock Springer, 2016.

\bibitem{rahwan2019machine}
Iyad Rahwan, Manuel Cebrian, Nick Obradovich, Josh Bongard, Jean-Fran{\c{c}}ois Bonnefon, Cynthia Breazeal, Jacob~W Crandall, Nicholas~A Christakis, Iain~D Couzin, Matthew~O Jackson, et~al.
\newblock Machine behaviour.
\newblock {\em Nature}, 568(7753):477--486, 2019.

\bibitem{johnson2013abrupt}
Neil Johnson, Guannan Zhao, Eric Hunsader, Hong Qi, Nicholas Johnson, Jing Meng, and Brian Tivnan.
\newblock Abrupt rise of new machine ecology beyond human response time.
\newblock {\em Scientific reports}, 3(1):2627, 2013.

\bibitem{aharony2011social}
Nadav Aharony, Wei Pan, Cory Ip, Inas Khayal, and Alex Pentland.
\newblock Social fmri: Investigating and shaping social mechanisms in the real world.
\newblock {\em Pervasive and mobile computing}, 7(6):643--659, 2011.

\bibitem{tinbergen2005aims}
Niko Tinbergen.
\newblock On aims and methods of ethology.
\newblock {\em Animal Biology}, 55(4), 2005.

\bibitem{burton2024large}
Jason~W Burton, Ezequiel Lopez-Lopez, Shahar Hechtlinger, Zoe Rahwan, Samuel Aeschbach, Michiel~A Bakker, Joshua~A Becker, Aleks Berditchevskaia, Julian Berger, Levin Brinkmann, et~al.
\newblock How large language models can reshape collective intelligence.
\newblock {\em Nature human behaviour}, pages 1--13, 2024.

\bibitem{crandall2018cooperating}
Jacob~W Crandall, Mayada Oudah, Tennom, Fatimah Ishowo-Oloko, Sherief Abdallah, Jean-Fran{\c{c}}ois Bonnefon, Manuel Cebrian, Azim Shariff, Michael~A Goodrich, and Iyad Rahwan.
\newblock Cooperating with machines.
\newblock {\em Nature communications}, 9(1):233, 2018.

\bibitem{sally1995conversation}
David Sally.
\newblock Conversation and cooperation in social dilemmas: A meta-analysis of experiments from 1958 to 1992.
\newblock {\em Rationality and society}, 7(1):58--92, 1995.

\bibitem{klein1993recognition}
Gary~A Klein.
\newblock A recognition-primed decision (rpd) model of rapid decision making.
\newblock {\em Decision making in action: Models and methods}, 5(4):138--147, 1993.

\bibitem{meta2022human}
Meta Fundamental AI Research Diplomacy~Team (FAIR)†, Anton Bakhtin, Noam Brown, Emily Dinan, Gabriele Farina, Colin Flaherty, Daniel Fried, Andrew Goff, Jonathan Gray, Hengyuan Hu, et~al.
\newblock Human-level play in the game of diplomacy by combining language models with strategic reasoning.
\newblock {\em Science}, 378(6624):1067--1074, 2022.

\bibitem{gent2023machine}
Edd Gent.
\newblock Machine mind readers.
\newblock {\em New Scientist}, 257(3426):46--49, 2023.

\bibitem{gigerenzer2024psychological}
Gerd Gigerenzer.
\newblock Psychological ai: Designing algorithms informed by human psychology.
\newblock {\em Perspectives on Psychological Science}, 19(5):839--848, 2024.

\bibitem{cui2024ai}
Hao Cui and Taha Yasseri.
\newblock Ai-enhanced collective intelligence.
\newblock {\em Patterns}, 5(11), 2024.

\bibitem{tsvetkova2024new}
Milena Tsvetkova, Taha Yasseri, Niccolo Pescetelli, and Tobias Werner.
\newblock A new sociology of humans and machines.
\newblock {\em Nature Human Behaviour}, 8(10):1864--1876, 2024.

\bibitem{kanwal2017comparing}
Maxinder~S Kanwal, Joshua~A Grochow, and Nihat Ay.
\newblock Comparing information-theoretic measures of complexity in boltzmann machines.
\newblock {\em Entropy}, 19(7):310, 2017.

\bibitem{watson2023collective}
Richard Watson and Michael Levin.
\newblock The collective intelligence of evolution and development.
\newblock {\em Collective Intelligence}, 2(2):26339137231168355, 2023.

\bibitem{frith2010social}
Uta Frith and Chris Frith.
\newblock The social brain: allowing humans to boldly go where no other species has been.
\newblock {\em Philosophical Transactions of the Royal Society B: Biological Sciences}, 365(1537):165--176, 2010.

\bibitem{penn2007lack}
Derek~C Penn and Daniel~J Povinelli.
\newblock On the lack of evidence that non-human animals possess anything remotely resembling a ‘theory of mind’.
\newblock {\em Philosophical Transactions of the Royal Society B: Biological Sciences}, 362(1480):731--744, 2007.

\bibitem{krupenye2019theory}
Christopher Krupenye and Josep Call.
\newblock Theory of mind in animals: Current and future directions.
\newblock {\em Wiley Interdisciplinary Reviews: Cognitive Science}, 10(6):e1503, 2019.

\bibitem{woolley2010evidence}
Anita~Williams Woolley, Christopher~F Chabris, Alex Pentland, Nada Hashmi, and Thomas~W Malone.
\newblock Evidence for a collective intelligence factor in the performance of human groups.
\newblock {\em science}, 330(6004):686--688, 2010.

\bibitem{engel2014reading}
David Engel, Anita~Williams Woolley, Lisa~X Jing, Christopher~F Chabris, and Thomas~W Malone.
\newblock Reading the mind in the eyes or reading between the lines? theory of mind predicts collective intelligence equally well online and face-to-face.
\newblock {\em PloS one}, 9(12):e115212, 2014.

\bibitem{yoshida2008game}
Wako Yoshida, Ray~J Dolan, and Karl~J Friston.
\newblock Game theory of mind.
\newblock {\em PLoS computational biology}, 4(12):e1000254, 2008.

\bibitem{gintis2006foundations}
Herbert Gintis.
\newblock The foundations of behavior: The beliefs, preferences, and constraints model.
\newblock {\em Biological Theory}, 1:123--127, 2006.

\bibitem{gintis2007framework}
Herbert Gintis.
\newblock A framework for the unification of the behavioral sciences.
\newblock {\em Behavioral and brain sciences}, 30(1):1--16, 2007.

\bibitem{gintis2007unifying}
Herbert Gintis.
\newblock Unifying the behavioral sciences ii.
\newblock {\em Behavioral and brain sciences}, 30(1):45--53, 2007.

\bibitem{pedreschi2024human}
Dino Pedreschi, Luca Pappalardo, Emanuele Ferragina, Ricardo Baeza-Yates, Albert-L{\'a}szl{\'o} Barab{\'a}si, Frank Dignum, Virginia Dignum, Tina Eliassi-Rad, Fosca Giannotti, J{\'a}nos Kert{\'e}sz, et~al.
\newblock Human-ai coevolution.
\newblock {\em Artificial Intelligence}, page 104244, 2024.

\bibitem{veissiere2020thinking}
Samuel~PL Veissi{\`e}re, Axel Constant, Maxwell~JD Ramstead, Karl~J Friston, and Laurence~J Kirmayer.
\newblock Thinking through other minds: A variational approach to cognition and culture.
\newblock {\em Behavioral and brain sciences}, 43:e90, 2020.

\bibitem{friston2024designing}
Karl~J Friston, Maxwell~JD Ramstead, Alex~B Kiefer, Alexander Tschantz, Christopher~L Buckley, Mahault Albarracin, Riddhi~J Pitliya, Conor Heins, Brennan Klein, Beren Millidge, et~al.
\newblock Designing ecosystems of intelligence from first principles.
\newblock {\em Collective Intelligence}, 3(1):26339137231222481, 2024.

\bibitem{harre2023testing}
Michael~S Harr{\'e} and Husam El-Tarifi.
\newblock Testing game theory of mind models for artificial intelligence.
\newblock {\em Games}, 15(1):1, 2023.

\bibitem{harre2022can}
Michael~S Harr{\'e}.
\newblock {What can game theory tell us about an AI ‘Theory of Mind’?}
\newblock {\em Games}, 13(3):46, 2022.

\bibitem{WoodRachel2019Iroe}
Rachel Wood, Alexander~G Liu, Frederick Bowyer, Philip~R Wilby, Frances~S Dunn, Charlotte~G Kenchington, Jennifer F~Hoyal Cuthill, Emily~G Mitchell, and Amelia Penny.
\newblock Integrated records of environmental change and evolution challenge the cambrian explosion.
\newblock {\em Nature ecology and evolution}, 3(4):528--538, 2019.

\bibitem{BaluškaFrantišek2016OHNH}
František Baluška and Michael Levin.
\newblock On having no head: Cognition throughout biological systems.
\newblock 7, 2016.

\bibitem{BoussardA2019Miap}
A~Boussard, J~Delescluse, A~Pérez-Escudero, and A~Dussutour.
\newblock Memory inception and preservation in slime moulds: the quest for a common mechanism.
\newblock {\em Philosophical transactions of the Royal Society of London. Series B. Biological sciences}, 374(1774):20180368--20180368, 2019.

\bibitem{ThompsonRichardF.2009HAh}
Richard~F. Thompson.
\newblock Habituation: A history.
\newblock {\em Neurobiology of learning and memory}, 92(2):127--134, 2009.

\bibitem{LiZhi2012QsHb}
Zhi Li and Satish~K. Nair.
\newblock Quorum sensing: How bacteria can coordinate activity and synchronize their response to external signals?
\newblock {\em Protein science}, 21(10):1403--1417, 2012.

\bibitem{avena2015network}
Andrea Avena-Koenigsberger, Joaqu{\'\i}n Go{\~n}i, Ricard Sol{\'e}, and Olaf Sporns.
\newblock Network morphospace.
\newblock {\em Journal of the Royal Society Interface}, 12(103):20140881, 2015.

\bibitem{claverie2014morphospace}
Thomas Claverie and Peter~C Wainwright.
\newblock A morphospace for reef fishes: elongation is the dominant axis of body shape evolution.
\newblock {\em PloS one}, 9(11):e112732, 2014.

\bibitem{seoane2018morphospace}
Lu{\'\i}s~F Seoane and Ricard Sol{\'e}.
\newblock The morphospace of language networks.
\newblock {\em Scientific reports}, 8(1):10465, 2018.

\bibitem{levin2022collective}
Michael Levin.
\newblock Collective intelligence of morphogenesis as a teleonomic process.
\newblock 2022.

\bibitem{Watson2011GlobalAdaptation}
Richard~A. Watson, Rob Mills, and C.~L. Buckley.
\newblock Global {{Adaptation}} in {{Networks}} of {{Selfish Components}}: {{Emergent Associative Memory}} at the {{System Scale}}.
\newblock {\em Artificial Life}, 17(3):147--166, July 2011.

\bibitem{Watson2011OptimizationSelfmodeling}
Richard~A. Watson, C.~L. Buckley, and Rob Mills.
\newblock Optimization in ``self-modeling'' complex adaptive systems.
\newblock {\em Complexity}, 16(5):17--26, 2011.

\bibitem{brede2022sensing}
Markus Brede and Guillermo Romero-Moreno.
\newblock Sensing enhancement on social networks: The role of network topology.
\newblock {\em Entropy}, 24(5):738, 2022.

\bibitem{KaoAlbertB2019Mswg}
Albert~B Kao and Iain~D Couzin.
\newblock Modular structure within groups causes information loss but can improve decision accuracy.
\newblock {\em Philosophical transactions of the Royal Society of London. Series B. Biological sciences}, 374(1774):20180378--20180378, 2019.

\bibitem{budd2021morphospace}
Graham~E Budd.
\newblock Morphospace.
\newblock {\em Current Biology}, 31(19):R1181--R1185, 2021.

\bibitem{arsiwalla2017morphospace}
Xerxes~D Arsiwalla, Ricard Sole, Clement Moulin-Frier, Ivan Herreros, Marti Sanchez-Fibla, and Paul Verschure.
\newblock The morphospace of consciousness.
\newblock {\em arXiv preprint arXiv:1705.11190}, 2017.

\bibitem{bahri2020statistical}
Yasaman Bahri, Jonathan Kadmon, Jeffrey Pennington, Sam~S Schoenholz, Jascha Sohl-Dickstein, and Surya Ganguli.
\newblock Statistical mechanics of deep learning.
\newblock {\em Annual Review of Condensed Matter Physics}, 11(1):501--528, 2020.

\bibitem{olle2016morphospace}
Aina Oll{\'e}-Vila, Salva Duran-Nebreda, N{\'u}ria Conde-Pueyo, Ra{\'u}l Monta{\~n}ez, and Ricard Sol{\'e}.
\newblock A morphospace for synthetic organs and organoids: the possible and the actual.
\newblock {\em Integrative Biology}, 8(4):485--503, 2016.

\bibitem{peterson2021using}
Joshua~C Peterson, David~D Bourgin, Mayank Agrawal, Daniel Reichman, and Thomas~L Griffiths.
\newblock Using large-scale experiments and machine learning to discover theories of human decision-making.
\newblock {\em Science}, 372(6547):1209--1214, 2021.

\bibitem{jara2019theory}
Julian Jara-Ettinger.
\newblock Theory of mind as inverse reinforcement learning.
\newblock {\em Current Opinion in Behavioral Sciences}, 29:105--110, 2019.

\bibitem{ruiz2023inverse}
Jaime Ruiz-Serra and Michael~S Harr{\'e}.
\newblock Inverse reinforcement learning as the algorithmic basis for theory of mind: current methods and open problems.
\newblock {\em Algorithms}, 16(2):68, 2023.

\bibitem{nguyen2022theory}
Thuy~Ngoc Nguyen and Cleotilde Gonzalez.
\newblock Theory of mind from observation in cognitive models and humans.
\newblock {\em Topics in cognitive science}, 14(4):665--686, 2022.

\bibitem{zhao2023teaching}
Michelle Zhao, Fade~R Eadeh, Thuy-Ngoc Nguyen, Pranav Gupta, Henny Admoni, Cleotilde Gonzalez, and Anita~Williams Woolley.
\newblock Teaching agents to understand teamwork: Evaluating and predicting collective intelligence as a latent variable via hidden markov models.
\newblock {\em Computers in Human Behavior}, 139:107524, 2023.

\bibitem{gupta2023fostering}
Pranav Gupta, Thuy~Ngoc Nguyen, Cleotilde Gonzalez, and Anita~Williams Woolley.
\newblock Fostering collective intelligence in human--ai collaboration: laying the groundwork for cohumain.
\newblock {\em Topics in cognitive science}, 2023.

\bibitem{shteynberg2023theory}
Garriy Shteynberg, Jacob~B Hirsh, Wouter Wolf, John~A Bargh, Erica~J Boothby, Andrew~M Colman, Gerald Echterhoff, and Maya Rossignac-Milon.
\newblock Theory of collective mind.
\newblock {\em Trends in Cognitive Sciences}, 27(11):1019--1031, 2023.

\bibitem{shum2019theory}
Michael Shum, Max Kleiman-Weiner, Michael~L Littman, and Joshua~B Tenenbaum.
\newblock Theory of minds: Understanding behavior in groups through inverse planning.
\newblock In {\em Proceedings of the AAAI conference on artificial intelligence}, volume~33, pages 6163--6170, 2019.

\bibitem{lombard2023causal}
Marlize Lombard and Peter G{\"a}rdenfors.
\newblock Causal cognition and theory of mind in evolutionary cognitive archaeology.
\newblock {\em Biological Theory}, 18(4):234--252, 2023.

\bibitem{haidle2010working}
Miriam~No{\"e}l Haidle.
\newblock Working-memory capacity and the evolution of modern cognitive potential: implications from animal and early human tool use.
\newblock {\em Current anthropology}, 51(S1):S149--S166, 2010.

\bibitem{lang2012pleistocene}
J{\"o}rg Lang, Jutta Winsemann, Dominik Steinmetz, Ulrich Polom, Lukas Pollok, Utz B{\"o}hner, Jordi Serangeli, Christian Brandes, Andrea Hampel, and Stefan Winghart.
\newblock The pleistocene of sch{\"o}ningen, germany: a complex tunnel valley fill revealed from 3d subsurface modelling and shear wave seismics.
\newblock {\em Quaternary Science Reviews}, 39:86--105, 2012.

\bibitem{happe2003theory}
Francesca Happ{\'e}.
\newblock Theory of mind and the self.
\newblock {\em Annals of the New York Academy of Sciences}, 1001(1):134--144, 2003.

\bibitem{birch2020dimensions}
Jonathan Birch, Alexandra~K Schnell, and Nicola~S Clayton.
\newblock Dimensions of animal consciousness.
\newblock {\em Trends in cognitive sciences}, 24(10):789--801, 2020.

\bibitem{chomsky2007biolinguistic}
Noam Chomsky.
\newblock Biolinguistic explorations: Design, development, evolution.
\newblock {\em International Journal of Philosophical Studies}, 15(1):1--21, 2007.

\bibitem{farrar2002early}
M~Jeffrey Farrar and Lisa Maag.
\newblock Early language development and the emergence of a theory of mind.
\newblock {\em First language}, 22(2):197--213, 2002.

\bibitem{de2021role}
Jill~G de~Villiers.
\newblock The role (s) of language in theory of mind.
\newblock In {\em The neural basis of mentalizing}, pages 423--448. Springer, 2021.

\bibitem{de2002complements}
Jill~G De~Villiers and Jennie~E Pyers.
\newblock Complements to cognition: A longitudinal study of the relationship between complex syntax and false-belief-understanding.
\newblock {\em Cognitive development}, 17(1):1037--1060, 2002.

\bibitem{milligan2007language}
Karen Milligan, Janet~Wilde Astington, and Lisa~Ain Dack.
\newblock Language and theory of mind: Meta-analysis of the relation between language ability and false-belief understanding.
\newblock {\em Child development}, 78(2):622--646, 2007.

\bibitem{thibodeau2011metaphors}
Paul~H Thibodeau and Lera Boroditsky.
\newblock Metaphors we think with: The role of metaphor in reasoning.
\newblock {\em PloS one}, 6(2):e16782, 2011.

\bibitem{thibodeau2017linguistic}
Paul~H Thibodeau, Rose~K Hendricks, and Lera Boroditsky.
\newblock How linguistic metaphor scaffolds reasoning.
\newblock {\em Trends in cognitive sciences}, 21(11):852--863, 2017.

\bibitem{bakshy2012role}
Eytan Bakshy, Itamar Rosenn, Cameron Marlow, and Lada Adamic.
\newblock The role of social networks in information diffusion.
\newblock In {\em Proceedings of the 21st international conference on World Wide Web}, pages 519--528, 2012.

\bibitem{dunbar1993coevolution}
Robin~IM Dunbar.
\newblock Coevolution of neocortical size, group size and language in humans.
\newblock {\em Behavioral and brain sciences}, 16(4):681--694, 1993.

\bibitem{dunbar2007evolution}
Robin~IM Dunbar and Susanne Shultz.
\newblock Evolution in the social brain.
\newblock {\em science}, 317(5843):1344--1347, 2007.

\bibitem{dunbar1995social}
Robin~IM Dunbar and Matt Spoors.
\newblock Social networks, support cliques, and kinship.
\newblock {\em Human nature}, 6:273--290, 1995.

\bibitem{lau2018discovering}
Tatiana Lau, Hillard~T Pouncy, Samuel~J Gershman, and Mina Cikara.
\newblock Discovering social groups via latent structure learning.
\newblock {\em Journal of Experimental Psychology: General}, 147(12):1881, 2018.

\bibitem{baron1997another}
Simon Baron-Cohen, Therese Jolliffe, Catherine Mortimore, and Mary Robertson.
\newblock Another advanced test of theory of mind: Evidence from very high functioning adults with autism or asperger syndrome.
\newblock {\em Journal of Child psychology and Psychiatry}, 38(7):813--822, 1997.

\bibitem{baron2001reading}
Simon Baron-Cohen, Sally Wheelwright, Jacqueline Hill, Yogini Raste, and Ian Plumb.
\newblock The “reading the mind in the eyes” test revised version: a study with normal adults, and adults with asperger syndrome or high-functioning autism.
\newblock {\em The Journal of Child Psychology and Psychiatry and Allied Disciplines}, 42(2):241--251, 2001.

\bibitem{harre2018strategic}
Michael~S Harr{\'e}.
\newblock Strategic information processing from behavioural data in iterated games.
\newblock {\em Entropy}, 20(1):27, 2018.

\bibitem{lee2004reinforcement}
Daeyeol Lee, Michelle~L Conroy, Benjamin~P McGreevy, and Dominic~J Barraclough.
\newblock Reinforcement learning and decision making in monkeys during a competitive game.
\newblock {\em Cognitive brain research}, 22(1):45--58, 2004.

\bibitem{bettencourt2009rules}
Lu{\'\i}s~MA Bettencourt.
\newblock The rules of information aggregation and emergence of collective intelligent behavior.
\newblock {\em Topics in Cognitive Science}, 1(4):598--620, 2009.

\bibitem{lizier2014jidt}
Joseph~T Lizier.
\newblock Jidt: An information-theoretic toolkit for studying the dynamics of complex systems.
\newblock {\em Frontiers in Robotics and AI}, 1:11, 2014.

\bibitem{wei2007plants}
Jianing Wei, Lizhong Wang, Junwei Zhu, Sufang Zhang, Owi~I Nandi, and Le~Kang.
\newblock Plants attract parasitic wasps to defend themselves against insect pests by releasing hexenol.
\newblock {\em PLOS one}, 2(9):e852, 2007.

\bibitem{kobayashi2007evolution}
Yutaka Kobayashi and Norio Yamamura.
\newblock Evolution of signal emission by uninfested plants to help nearby infested relatives.
\newblock {\em Evolutionary Ecology}, 21:281--294, 2007.

\bibitem{golubski2016ecological}
Antonio~J Golubski, Erik~E Westlund, John Vandermeer, and Mercedes Pascual.
\newblock Ecological networks over the edge: hypergraph {Trait-Mediated Indirect Interaction (TMII)} structure.
\newblock {\em Trends in ecology \& evolution}, 31(5):344--354, 2016.

\bibitem{harris2023smooth}
Adam Harris, Scott McCallum, and Michael~S Harr{\'e}.
\newblock On the smooth unfolding of bifurcations in quantal-response equilibria.
\newblock {\em Games and Economic Behavior}, 2023.

\bibitem{harre2018multi}
Michael~S Harr{\'e}.
\newblock Multi-agent economics and the emergence of critical markets.
\newblock {\em arXiv preprint arXiv:1809.01332}, 2018.

\bibitem{friston2023free}
Karl Friston, Lancelot Da~Costa, Noor Sajid, Conor Heins, Kai Ueltzh{\"o}ffer, Grigorios~A Pavliotis, and Thomas Parr.
\newblock The free energy principle made simpler but not too simple.
\newblock {\em Physics Reports}, 1024:1--29, 2023.

\bibitem{ruiz2024factorised}
Jaime Ruiz-Serra, Patrick Sweeney, and Michael~S Harr{\'e}.
\newblock Factorised active inference for strategic multi-agent interactions.
\newblock {\em arXiv preprint arXiv:2411.07362}, 2024.

\bibitem{clark2020niche}
Andrew~D Clark, Dominik Deffner, Kevin Laland, John Odling-Smee, and John Endler.
\newblock Niche construction affects the variability and strength of natural selection.
\newblock {\em The American Naturalist}, 195(1):16--30, 2020.

\bibitem{trappes2022individualized}
Rose Trappes, Behzad Nematipour, Marie~I Kaiser, Ulrich Krohs, Koen~J Van~Benthem, Ulrich~R Ernst, J{\"u}rgen Gadau, Peter Korsten, Joachim Kurtz, Holger Schielzeth, et~al.
\newblock How individualized niches arise: Defining mechanisms of niche construction, niche choice, and niche conformance.
\newblock {\em BioScience}, 72(6):538--548, 2022.

\bibitem{schmalzle2017brain}
Ralf Schm{\"a}lzle, Matthew Brook~O’Donnell, Javier~O Garcia, Christopher~N Cascio, Joseph Bayer, Danielle~S Bassett, Jean~M Vettel, and Emily~B Falk.
\newblock Brain connectivity dynamics during social interaction reflect social network structure.
\newblock {\em Proceedings of the National Academy of Sciences}, 114(20):5153--5158, 2017.

\bibitem{campbell2016universal}
John~O Campbell.
\newblock Universal darwinism as a process of bayesian inference.
\newblock {\em Frontiers in systems neuroscience}, 10:49, 2016.

\bibitem{watson2016evolutionary}
Richard~A Watson, Rob Mills, CL~Buckley, Kostas Kouvaris, Adam Jackson, Simon~T Powers, Chris Cox, Simon Tudge, Adam Davies, Loizos Kounios, et~al.
\newblock Evolutionary connectionism: algorithmic principles underlying the evolution of biological organisation in evo-devo, evo-eco and evolutionary transitions.
\newblock {\em Evolutionary biology}, 43:553--581, 2016.

\bibitem{watson2016can}
Richard~A Watson and E{\"o}rs Szathm{\'a}ry.
\newblock How can evolution learn?
\newblock {\em Trends in ecology \& evolution}, 31(2):147--157, 2016.

\bibitem{phelps2010evolutionary}
Steve Phelps, Peter McBurney, and Simon Parsons.
\newblock Evolutionary mechanism design: a review.
\newblock {\em Autonomous agents and multi-agent systems}, 21(2):237--264, 2010.

\bibitem{McKee2023ScaffoldingCooperation}
Kevin~R. McKee, Andrea Tacchetti, Michiel~A. Bakker, Jan Balaguer, Lucy {Campbell-Gillingham}, Richard Everett, and Matthew Botvinick.
\newblock Scaffolding cooperation in human groups with deep reinforcement learning.
\newblock {\em Nature Human Behaviour}, 7(10):1787--1796, October 2023.

\bibitem{Westby2023CollectiveIntelligence}
Samuel Westby and Christoph Riedl.
\newblock Collective {{Intelligence}} in {{Human-AI Teams}}: {{A Bayesian Theory}} of {{Mind Approach}}.
\newblock {\em Proceedings of the AAAI Conference on Artificial Intelligence}, 37(5):6119--6127, June 2023.

\bibitem{Bendell2024IndividualTeam}
Rhyse Bendell, Jessica Williams, Stephen~M. Fiore, and Florian Jentsch.
\newblock Individual and team profiling to support theory of mind in artificial social intelligence.
\newblock {\em Scientific Reports}, 14(1):12635, June 2024.

\bibitem{Sarkadi2023ArmsRace}
{\c S}tefan Sarkadi.
\newblock An {{Arms Race}} in {{Theory-of-Mind}}: {{Deception Drives}} the {{Emergence}} of {{Higher-level Theory-of-Mind}} in {{Agent Societies}}.
\newblock In {\em 2023 {{IEEE International Conference}} on {{Autonomic Computing}} and {{Self-Organizing Systems}} ({{ACSOS}})}, pages 1--10, September 2023.

\bibitem{Bao2024ReadingMarket}
Te~Bao, Sascha F{\"u}llbrunn, Jiaoying Pei, and Jichuan Zong.
\newblock Reading the market? {{Expectation}} coordination and theory of mind.
\newblock {\em Journal of Economic Behavior \& Organization}, 219:510--527, March 2024.

\bibitem{Garcia-Lopez2023TheoryMind}
Alvaro {Garcia-Lopez}.
\newblock Theory of {{Mind}} in {{Artificial Intelligence Applications}}.
\newblock In Teresa {Lopez-Soto}, Alvaro {Garcia-Lopez}, and Francisco~J. {Salguero-Lamillar}, editors, {\em The {{Theory}} of {{Mind Under Scrutiny}}: {{Psychopathology}}, {{Neuroscience}}, {{Philosophy}} of {{Mind}} and {{Artificial Intelligence}}}, pages 723--750. Springer Nature Switzerland, Cham, 2023.

\bibitem{Rocha2023ApplyingTheory}
Michele Rocha, Heitor~Henrique {da Silva}, Anal{\'u}cia~Schiaffino Morales, Stefan Sarkadi, and Alison~R. Panisson.
\newblock Applying {{Theory}} of~{{Mind}} to~{{Multi-agent Systems}}: {{A Systematic Review}}.
\newblock In Murilo~C. Naldi and Reinaldo A.~C. Bianchi, editors, {\em Intelligent {{Systems}}}, pages 367--381, Cham, 2023. Springer Nature Switzerland.

\bibitem{Nebreda2023SocialMachine}
Alberto Nebreda, Danylyna {Shpakivska-Bilan}, Carmen Camara, and Gianluca Susi.
\newblock The {{Social Machine}}: {{Artificial Intelligence}} ({{AI}}) {{Approaches}} to {{Theory}} of {{Mind}}.
\newblock In Teresa {Lopez-Soto}, Alvaro {Garcia-Lopez}, and Francisco~J. {Salguero-Lamillar}, editors, {\em The {{Theory}} of {{Mind Under Scrutiny}}: {{Psychopathology}}, {{Neuroscience}}, {{Philosophy}} of {{Mind}} and {{Artificial Intelligence}}}, pages 681--722. Springer Nature Switzerland, Cham, 2023.

\bibitem{awad2018moral}
Edmond Awad, Sohan Dsouza, Richard Kim, Jonathan Schulz, Joseph Henrich, Azim Shariff, Jean-Fran{\c{c}}ois Bonnefon, and Iyad Rahwan.
\newblock The moral machine experiment.
\newblock {\em Nature}, 563(7729):59--64, 2018.

\bibitem{mao2024review}
Yuanyuan Mao, Shuang Liu, Qin Ni, Xin Lin, and Liang He.
\newblock A review on machine theory of mind.
\newblock {\em IEEE Transactions on Computational Social Systems}, 2024.

\bibitem{de2022understanding}
Javier Fern{\'a}ndez-L{\'o}pez de~Pablo, Val{\'e}ria Romano, Maxime Derex, Erik Gjesfjeld, Claudine Gravel-Miguel, Marcus~J Hamilton, Andrea~Bamberg Migliano, Felix Riede, and Sergi Lozano.
\newblock Understanding hunter--gatherer cultural evolution needs network thinking.
\newblock {\em Trends in Ecology \& Evolution}, 37(8):632--636, 2022.

\bibitem{dyble2016networks}
Mark Dyble, James Thompson, Daniel Smith, Gul~Deniz Salali, Nikhil Chaudhary, Abigail~E Page, Lucio Vinicuis, Ruth Mace, and Andrea~Bamberg Migliano.
\newblock Networks of food sharing reveal the functional significance of multilevel sociality in two hunter-gatherer groups.
\newblock {\em Current Biology}, 26(15):2017--2021, 2016.

\bibitem{laland2016introduction}
Kevin Laland, Blake Matthews, and Marcus~W Feldman.
\newblock An introduction to niche construction theory.
\newblock {\em Evolutionary ecology}, 30:191--202, 2016.

\bibitem{rowley2011foraging}
Peter Rowley-Conwy and Robert Layton.
\newblock Foraging and farming as niche construction: stable and unstable adaptations.
\newblock {\em Philosophical Transactions of the Royal Society B: Biological Sciences}, 366(1566):849--862, 2011.

\bibitem{arroyo2017civilisation}
Manuel Arroyo-Kalin, David Wengrow, Dorian~Q Fuller, Chris~J Stevens, and Mich{\`e}le Wollstonecroft.
\newblock Civilisation and human niche construction.
\newblock {\em Archaeology International}, 20(1):106--109, 2017.

\bibitem{kemp20207000}
Melissa~E Kemp, Alexis~M Mychajliw, Jenna Wadman, and Amy Goldberg.
\newblock 7000 years of turnover: historical contingency and human niche construction shape the caribbean's anthropocene biota.
\newblock {\em Proceedings of the Royal Society B}, 287(1927):20200447, 2020.

\bibitem{smith2013onset}
Bruce~D Smith and Melinda~A Zeder.
\newblock The onset of the anthropocene.
\newblock {\em Anthropocene}, 4:8--13, 2013.

\bibitem{ellis2024anthropocene}
Erle~C Ellis.
\newblock The anthropocene condition: evolving through social--ecological transformations.
\newblock {\em Philosophical Transactions of the Royal Society B}, 379(1893):20220255, 2024.

\bibitem{wilson2023multilevel}
David~Sloan Wilson, Guru Madhavan, Michele~J Gelfand, Steven~C Hayes, Paul~WB Atkins, and Rita~R Colwell.
\newblock Multilevel cultural evolution: From new theory to practical applications.
\newblock {\em Proceedings of the National Academy of Sciences}, 120(16):e2218222120, 2023.

\bibitem{ryan2016social}
Paul~A Ryan, Simon~T Powers, and Richard~A Watson.
\newblock Social niche construction and evolutionary transitions in individuality.
\newblock {\em Biology \& philosophy}, 31:59--79, 2016.

\bibitem{vaswani2017attention}
A~Vaswani.
\newblock Attention is all you need.
\newblock {\em Advances in Neural Information Processing Systems}, 2017.

\bibitem{silver2021reward}
David Silver, Satinder Singh, Doina Precup, and Richard~S Sutton.
\newblock Reward is enough.
\newblock {\em Artificial Intelligence}, 299:103535, 2021.

\bibitem{li2024chain}
Zhiyuan Li, Hong Liu, Denny Zhou, and Tengyu Ma.
\newblock Chain of thought empowers transformers to solve inherently serial problems.
\newblock {\em arXiv preprint arXiv:2402.12875}, 2024.

\bibitem{wei2022emergent}
Jason Wei, Yi~Tay, Rishi Bommasani, Colin Raffel, Barret Zoph, Sebastian Borgeaud, Dani Yogatama, Maarten Bosma, Denny Zhou, Donald Metzler, et~al.
\newblock Emergent abilities of large language models.
\newblock {\em arXiv preprint arXiv:2206.07682}, 2022.

\bibitem{strachan2024testing}
James~WA Strachan, Dalila Albergo, Giulia Borghini, Oriana Pansardi, Eugenio Scaliti, Saurabh Gupta, Krati Saxena, Alessandro Rufo, Stefano Panzeri, Guido Manzi, et~al.
\newblock Testing theory of mind in large language models and humans.
\newblock {\em Nature Human Behaviour}, pages 1--11, 2024.

\bibitem{bubeck2023sparks}
S{\'e}bastien Bubeck, Varun Chandrasekaran, Ronen Eldan, Johannes Gehrke, Eric Horvitz, Ece Kamar, Peter Lee, Yin~Tat Lee, Yuanzhi Li, Scott Lundberg, et~al.
\newblock Sparks of artificial general intelligence: Early experiments with gpt-4.
\newblock {\em arXiv preprint arXiv:2303.12712}, 2023.

\bibitem{xiong2022artificial}
Momiao Xiong.
\newblock {\em Artificial Intelligence and Causal Inference}.
\newblock Chapman and Hall/CRC, 2022.

\bibitem{laland2014does}
Kevin Laland, Tobias Uller, Marc Feldman, Kim Sterelny, Gerd~B M{\"u}ller, Armin Moczek, Eva Jablonka, John Odling-Smee, Gregory~A Wray, Hopi~E Hoekstra, et~al.
\newblock Does evolutionary theory need a rethink?
\newblock {\em Nature}, 514(7521):161--164, 2014.

\bibitem{szathmary1995major}
E{\"o}rs Szathm{\'a}ry and John~Maynard Smith.
\newblock The major evolutionary transitions.
\newblock {\em Nature}, 374(6519):227--232, 1995.

\bibitem{szathmary2015toward}
E{\"o}rs Szathm{\'a}ry.
\newblock Toward major evolutionary transitions theory 2.0.
\newblock {\em Proceedings of the National Academy of Sciences}, 112(33):10104--10111, 2015.

\bibitem{prokopenko2024biological}
Mikhail Prokopenko, Paul~CW Davies, Michael Harr{\'e}, Marcus Heisler, Zdenka Kuncic, Geraint~F Lewis, Ori Livson, Joseph~T Lizier, and Fernando~E Rosas.
\newblock Biological arrow of time: Emergence of tangled information hierarchies and self-modelling dynamics.
\newblock {\em arXiv preprint arXiv:2409.12029}, 2024.

\bibitem{wilson2024rethinking}
David~Sloan Wilson and Dennis~J Snower.
\newblock Rethinking the theoretical foundation of economics i: The multilevel paradigm.
\newblock {\em Economics}, 18(1):20220070, 2024.

\bibitem{czegel2019multilevel}
D{\'a}niel Cz{\'e}gel, Istv{\'a}n Zachar, and E{\"o}rs Szathm{\'a}ry.
\newblock Multilevel selection as bayesian inference, major transitions in individuality as structure learning.
\newblock {\em Royal Society open science}, 6(8):190202, 2019.

\end{thebibliography}


\appendix
\counterwithin*{equation}{section}
\renewcommand\theequation{\thesection\arabic{equation}}
\newpage

\section{Utility polynomials}\label{apx:utility-polynomials}

In two-player, $m$-action normal-form games, the payoffs are encoded in a single $m\times m$ matrix $\texttt{G}$, where each cell contains the payoffs for the $i$ and $j$ players for a particular combination of actions.

\begin{table}[h]
    \centering
    \begin{tabular}{|c|l|c|c|}
        \cline{3-4}
        \multicolumn{2}{c|}{} & \multicolumn{2}{c|}{$x_j$} \\ \cline{3-4}
        \multicolumn{2}{c|}{} & $\texttt{C}$ & $\texttt{D}$ \\ \hline
        \multirow{2}{*}{$x_i$} & $\texttt{C}$ & $(g_i^{cc},g_j^{cc})$ & $(g_i^{cd},g_j^{dc})$ \\ \cline{2-4}
                               & $\texttt{D}$ & $(g_i^{dc},g_j^{cd})$ & $(g_i^{dd},g_j^{dd})$ \\ \hline
    \end{tabular}
    \caption{Two-player, two-action normal-form game payoff matrix $\texttt{G}$}
    \label{tab:normal-form-matrix}
\end{table}

The payoff matrix $\texttt{G}$ can be decomposed into two standard matrices $\texttt{G}_i, \texttt{G}_j$ containing each player's respective payoffs.
Utility functions can be obtained in polynomial form from the elements $g_i^{l} \in \texttt{G}_i$, where each term intuitively represents the contribution of each agent's action (first-order terms) and their interaction (second-order terms) \cite{harre2018multi}. For $m=2$ actions, we have:
\begin{equation}
    U_i(\mathbf{x}; \mathbf{a}_i) =
    a_i^0 + a_i^i x_i + a_i^j x_j + a_i^{ij} x_i x_j
\end{equation}
\begin{equation} \label{eq:utility-polynomial}
    \mathbf{a}_i = 
    \begin{bmatrix} 
        a_i^0  \\ 
        a_i^i  \\ 
        a_i^j  \\ 
        a_i^{ij} 
    \end{bmatrix}
    =
    \frac{1}{4}
    \begin{bmatrix} 
        g_i^{cc} + g_i^{cd} + g_i^{dc} + g_i^{dd} \\
        (g_i^{dc} + g_i^{dd}) - (g_i^{cc} + g_i^{cd}) \\
        (g_i^{cd} + g_i^{dd}) - (g_i^{cc} + g_i^{dc}) \\
        (g_i^{cc} + g_i^{dd}) - (g_i^{cd} + g_i^{dc}) 
    \end{bmatrix}
\end{equation}

\subsection{General form for $n$ players and $m=2$ actions}

For any number $n$ of players where each player's action $x_i \in \{-1, 1\}$, the utility function is
\begin{equation}
    U_i(\mathbf{x}; \mathbf{a}_i) = \sum_{S \subseteq \{1, 2, \ldots, n\}} a_i^S \prod_{j \in S} x_j
\end{equation}
where $S$ is any subset of the set $\{1, 2, \ldots, n\}$, and $a_i^S$ is the co-factor corresponding to the subset $S$. 
For example, $S = \{1, 2, 3\}$ means the product $\prod_{j \in S} x_j = x_1 x_2 x_3$, and $a_i^S = a_i^{1,2,3}$ represents the co-factor for this interaction. 
To determine the co-factors $a_i^S$, we need to consider the payoff matrix $\texttt{G}_i$, which contains $2^n$ entries corresponding to each combination of $n$ players' actions. 
Denote the entry in $\texttt{G}_i$ for the action combination $(x_1, x_2, \ldots, x_n)$ as $g_i^{x_1 x_2 \ldots x_n}$.
The co-factors $a_i^S$ are:
\begin{equation}
    a_i^S = 
        \frac{1}{2^n} 
        \sum_{(x_1, x_2, \ldots, x_n) \in \{-1, 1\}^n} 
        g_i^{x_1 x_2 \ldots x_n} \prod_{j \in S} x_j
\end{equation}

This generalization for the utility function for $n$-player, two-action games can be interpreted through Fourier analysis on the Boolean cube. This connection arises because we represent the utility function as a sum of basis functions over the Boolean domain, where the co-factors $a_i^S$ are Fourier co-factors and $\prod_{j \in S} x_j$ are parity functions. Thus, we express the utility function $U_i$ as a Fourier transform of the payoff matrix $\texttt{G}_i$. Each co-factor $a_i^S$ captures the influence of a subset $S$ of players on the overall utility, analogous to how Fourier co-factors capture the influence of different frequency components in a signal. This generalization leverages the principles of Fourier analysis on the Boolean cube to decompose the complex payoff interactions into simpler, orthogonal components, making it easier to analyze and interpret the contributions of different action combinations in the game.

\newpage

\section{Derivation of utility polynomials in example}\label{apx:utility-derivation}

From the provided payoff matrix,
\begin{table}[h]
    \centering
    \begin{tabular}{|l|c|c|}
        \cline{2-3}
        \multicolumn{1}{c|}{} & \texttt{C} & \texttt{D} \\ \hline
        \texttt{C} & $(1,1)$ & $(0-c,1+c)$ \\ \hline
        \texttt{D} & $(1+c,0-c)$ & $(0,0)$ \\ \hline
    \end{tabular}
\end{table}

\noindent we can obtain the polynomial co-factors (see Table~\ref{tab:normal-form-matrix} and Equation~\ref{eq:utility-polynomial} for details)
\begin{align}
    \mathbf{a}_i = 
    \begin{bmatrix} 
        a_i^0  \\ 
        a_i^i  \\ 
        a_i^j  \\ 
        a_i^{ij} 
    \end{bmatrix}
    &=
    \frac{1}{4}
    \begin{bmatrix} 
        g_i^{cc} + g_i^{cd} + g_i^{dc} + g_i^{dd} \\
        (g_i^{dc} + g_i^{dd}) - (g_i^{cc} + g_i^{cd}) \\
        (g_i^{cd} + g_i^{dd}) - (g_i^{cc} + g_i^{dc}) \\
        (g_i^{cc} + g_i^{dd}) - (g_i^{cd} + g_i^{dc}) 
    \end{bmatrix} \\
    &=
    \frac{1}{4}
    \begin{bmatrix} 
        1 + (0-c) + (1+c) + 0 \\
        ((1+c) + 0) - (1 + (0-c)) \\
        ((0-c) + 0) - (1 + (1+c)) \\
        (1 + 0) - ((0-c) + (1+c)) 
    \end{bmatrix}
    =
    \frac{1}{4}
    \begin{bmatrix} 
        1 + 1 \\
        1 + c - 1  + c \\
        -c -1 -1 -c \\
        1 +c -1 -c 
    \end{bmatrix} \\
    &=
    \frac{1}{4}
    \begin{bmatrix} 
        2 \\
        2c \\
        -2(1+c) \\
        0
    \end{bmatrix}
    =
    \frac{1}{2}
    \begin{bmatrix} 
        1 \\
        c \\
        -(1+c) \\
        0
    \end{bmatrix}
    =
    \begin{bmatrix} 
        a_i^0  \\ 
        a_i^i  \\ 
        a_i^j  \\ 
        a_i^{ij} 
    \end{bmatrix}
\end{align}

\noindent and substitute into the (second-degree) utility polynomial (with zero quadratic term in this particular case)
\begin{equation}
    U_i(\mathbf{x};\mathbf{a})
        = a_i^0 + a_i^i x_i + a_i^j x_j + a_i^{ij} x_i x_j 
        = \frac{1}{2}\Big(1 + c x_i - (1+c) x_j \Big)
\end{equation}

\noindent we can choose to re-scale the utility values by a factor of 2, which results in the following utility polynomials for $A_2$ and $A_3$, respectively
\begin{align}
    U_2(\mathbf{x};\mathbf{a}) &= 1 + cx_2 - (1+c)x_3, &\,\,  
    U_3(\mathbf{x};\mathbf{a}) &= 1 + cx_3 - (1+c)x_2.
\end{align}

\end{document}